\documentclass[prd,twocolumn,amsmath,showpacs,amssymb,floatfix,superscriptaddress]{revtex4-1}

\usepackage{bm}
\usepackage{amsmath}
\usepackage{epsf}
\usepackage{color}
\usepackage{graphicx}
\usepackage{verbatim}
\usepackage[shortlabels]{enumitem}
\setlist[enumerate, 1]{1\textsuperscript{o}}

\usepackage{soul}

\newcommand{\bea}{\begin{eqnarray}}
\newcommand{\eea}{\end{eqnarray}}

\newcommand{\OmegaDE}{\Omega_{{\rm DE}}}
\newcommand{\Omegam}{\Omega_{{\rm m}}}

\newcommand{\Msun}{M_\odot}
\newcommand{\Mobs}{M^{\rm obs}}
\newcommand{\zphot}{z^{\rm phot}}

\newcommand{\zm}{{z_{max}}}
\newcommand{\rich}{M^{\rm obs}}

\newcommand{\fixed}[1]{{#1}}
\newcommand{\fixedII}[1]{{\bf #1}}

\definecolor{darkgreen}{cmyk}{0.85,0.2,1.00,0.2} 
\definecolor{purple}{cmyk}{0.5,1.0,0,0} 

\def\mnras{Mon.\ Not.\ R.\ Astron.\ Soc.\ }

\def\aj{Astron.~J.}
\newcommand{\apjs}{{Astrophys.~J.~Supp.}}
\newcommand{\jcap}{{Journal of Cosmology and Astroparticle Physics}}
\newcommand{\aap}{{Astron.~Astrophys.}}

\newcommand{\eqn}[1]{\bea#1\eea}

\newcommand{\de}[1]{\left( #1 \right)}

\newcommand{\cuad}[1]{\left[ #1 \right]}

\newcommand{\virg}[1]{_{,#1}}

\newcommand{\del}[2]{#1{\virg{#2}}}

\newcommand{\al}{\alpha}
\newcommand{\be}{\beta}

\newcommand{\mm}{{\bar m}}

\newcommand{\MM}{\pmb{\mm}}

\newcommand{\Tr}[1]{\text{Tr}\cuad{#1}}
\newcommand{\Cov}{\pmb{C}}

\newcommand{\Covs}{\pmb{S}}
\newcommand{\Covm}{\pmb{M}}
\newcommand{\Covi}{\Cov^{-1}}

\newcommand{\code}[1]{\texttt{#1}}

\newcommand{\select}{case_n500_m138_new}
\newcommand{\mth}{13.8}

\newcommand{\logMobs}[1]{\log[\Mobs_{#1}/(\Msun h^{-1})]}
\newcommand{\prv}{1}

\begin{document}
\newcommand{\sigxxOmA}{0.006}
\newcommand{\sigfxOmA}{0.010}
\newcommand{\sigxfOmA}{0.011}
\newcommand{\sigffOmA}{0.012}
\newcommand{\degffOmA}{105\%}
\newcommand{\sigpfOmA}{0.011}
\newcommand{\sigfpOmA}{0.010}
\newcommand{\sigppOmA}{0.006}
\newcommand{\degppOmA}{7\%}
\newcommand{\degfffxOmA}{27\%}

\newcommand{\sigxxwA}{0.035}
\newcommand{\sigfxwA}{0.046}
\newcommand{\sigxfwA}{0.045}
\newcommand{\sigffwA}{0.052}
\newcommand{\degffwA}{49\%}
\newcommand{\sigpfwA}{0.045}
\newcommand{\sigfpwA}{0.046}
\newcommand{\sigppwA}{0.041}
\newcommand{\degppwA}{18\%}
\newcommand{\degfffxwA}{12\%}

\newcommand{\sigxxOmB}{0.006}
\newcommand{\sigfxOmB}{0.009}
\newcommand{\sigxfOmB}{0.009}
\newcommand{\sigffOmB}{0.010}
\newcommand{\degffOmB}{70\%}
\newcommand{\sigpfOmB}{0.009}
\newcommand{\sigfpOmB}{0.009}
\newcommand{\sigppOmB}{0.006}
\newcommand{\degppOmB}{8\%}
\newcommand{\degfffxOmB}{4\%}

\newcommand{\sigxxwB}{0.033}
\newcommand{\sigfxwB}{0.044}
\newcommand{\sigxfwB}{0.042}
\newcommand{\sigffwB}{0.046}
\newcommand{\degffwB}{36\%}
\newcommand{\sigpfwB}{0.042}
\newcommand{\sigfpwB}{0.044}
\newcommand{\sigppwB}{0.041}
\newcommand{\degppwB}{21\%}
\newcommand{\degfffxwB}{2\%}

\newcommand{\sigxxOmC}{0.006}
\newcommand{\sigfxOmC}{0.010}
\newcommand{\sigxfOmC}{0.010}
\newcommand{\sigffOmC}{0.012}
\newcommand{\degffOmC}{86\%}
\newcommand{\sigpfOmC}{0.010}
\newcommand{\sigfpOmC}{0.010}
\newcommand{\sigppOmC}{0.007}
\newcommand{\degppOmC}{6\%}
\newcommand{\degfffxOmC}{15\%}

\newcommand{\sigxxwC}{0.036}
\newcommand{\sigfxwC}{0.047}
\newcommand{\sigxfwC}{0.045}
\newcommand{\sigffwC}{0.049}
\newcommand{\degffwC}{34\%}
\newcommand{\sigpfwC}{0.045}
\newcommand{\sigfpwC}{0.048}
\newcommand{\sigppwC}{0.042}
\newcommand{\degppwC}{16\%}
\newcommand{\degfffxwC}{3\%}

\newcommand{\sigfixfix}{(\sigxxOmB, \sigxxwB)}
\newcommand{\sigfrefix}{(\sigfxOmB, \sigfxwB)}
\newcommand{\sigfixfre}{(\sigxfOmB, \sigxfwB)}
\newcommand{\sigfrefre}{(\sigffOmB, \sigffwB)}
\newcommand{\sigprifre}{(\sigpfOmB, \sigpfwB)}
\newcommand{\sigfrepri}{(\sigfpOmB, \sigfpwB)}
\newcommand{\sigpripri}{(\sigppOmB, \sigppwB)}
\newcommand{\degfrefre}{(\degffOmB, \degffwB)}
\newcommand{\degpripri}{(\degppOmB, \degppwB)}
\newcommand{\degfrefrefx}{(\degfffxOmB, \degfffxwB)}

\newcommand{\sigxxOmZ}{0.006}
\newcommand{\sigfxOmZ}{0.010}
\newcommand{\sigxfOmZ}{0.006}
\newcommand{\sigxxwZ}{0.034}
\newcommand{\sigfxwZ}{0.048}
\newcommand{\sigxfwZ}{0.038}

\newcommand{\sigfrenul}{(\sigfxOmZ, \sigfxwZ)}

\newcommand{\sigOmAIII     }{0.086}	 \newcommand{\degOmAIII     }{597\%}
\newcommand{\sigOmAIV      }{0.048}	 \newcommand{\degOmAIV      }{286\%}
\newcommand{\sigOmAV       }{0.029}	 \newcommand{\degOmAV       }{133\%}
\newcommand{\sigOmAVI      }{0.021}	 \newcommand{\degOmAVI      }{71\%}
\newcommand{\sigOmAVII     }{0.018}	 \newcommand{\degOmAVII     }{49\%}
\newcommand{\sigOmAVIII    }{0.016}	 \newcommand{\degOmAVIII    }{25\%}
\newcommand{\sigOmAIX      }{0.014}	 \newcommand{\degOmAIX      }{11\%}
\newcommand{\sigOmAX       }{0.012}	 \newcommand{\degOmAX       }{0\%}
\newcommand{\sigOmAXI      }{0.011}	 \newcommand{\degOmAXI      }{7\%}
\newcommand{\sigOmAXII     }{0.011}	 \newcommand{\degOmAXII     }{14\%}
\newcommand{\sigOmAXIII    }{0.010}	 \newcommand{\degOmAXIII    }{18\%}
\newcommand{\sigOmAXIV     }{0.010}	 \newcommand{\degOmAXIV     }{19\%}
\newcommand{\sigOmAXV      }{0.010}	 \newcommand{\degOmAXV      }{22\%}
\newcommand{\sigOmAXVI     }{0.009}	 \newcommand{\degOmAXVI     }{24\%}
\newcommand{\sigOmAXVII    }{0.009}	 \newcommand{\degOmAXVII    }{27\%}
\newcommand{\sigOmAXVIII   }{0.009}	 \newcommand{\degOmAXVIII   }{27\%}
\newcommand{\sigOmAXIX     }{0.009}	 \newcommand{\degOmAXIX     }{28\%}
\newcommand{\sigOmAXX      }{0.009}	 \newcommand{\degOmAXX      }{28\%}

\newcommand{\sigwAIII     }{0.247}	 \newcommand{\degwAIII     }{374\%}
\newcommand{\sigwAIV      }{0.176}	 \newcommand{\degwAIV      }{237\%}
\newcommand{\sigwAV       }{0.105}	 \newcommand{\degwAV       }{102\%}
\newcommand{\sigwAVI      }{0.087}	 \newcommand{\degwAVI      }{66\%}
\newcommand{\sigwAVII     }{0.081}	 \newcommand{\degwAVII     }{56\%}
\newcommand{\sigwAVIII    }{0.066}	 \newcommand{\degwAVIII    }{27\%}
\newcommand{\sigwAIX      }{0.058}	 \newcommand{\degwAIX      }{10\%}
\newcommand{\sigwAX       }{0.052}	 \newcommand{\degwAX       }{0\%}
\newcommand{\sigwAXI      }{0.049}	 \newcommand{\degwAXI      }{5\%}
\newcommand{\sigwAXII     }{0.047}	 \newcommand{\degwAXII     }{9\%}
\newcommand{\sigwAXIII    }{0.045}	 \newcommand{\degwAXIII    }{13\%}
\newcommand{\sigwAXIV     }{0.041}	 \newcommand{\degwAXIV     }{20\%}
\newcommand{\sigwAXV      }{0.040}	 \newcommand{\degwAXV      }{23\%}
\newcommand{\sigwAXVI     }{0.039}	 \newcommand{\degwAXVI     }{24\%}
\newcommand{\sigwAXVII    }{0.038}	 \newcommand{\degwAXVII    }{26\%}
\newcommand{\sigwAXVIII   }{0.037}	 \newcommand{\degwAXVIII   }{28\%}
\newcommand{\sigwAXIX     }{0.036}	 \newcommand{\degwAXIX     }{30\%}
\newcommand{\sigwAXX      }{0.035}	 \newcommand{\degwAXX      }{32\%}

\newcommand{\sigOmBIII     }{0.033}	 \newcommand{\degOmBIII     }{235\%}
\newcommand{\sigOmBIV      }{0.024}	 \newcommand{\degOmBIV      }{141\%}
\newcommand{\sigOmBV       }{0.018}	 \newcommand{\degOmBV       }{88\%}
\newcommand{\sigOmBVI      }{0.015}	 \newcommand{\degOmBVI      }{58\%}
\newcommand{\sigOmBVII     }{0.014}	 \newcommand{\degOmBVII     }{43\%}
\newcommand{\sigOmBVIII    }{0.012}	 \newcommand{\degOmBVIII    }{23\%}
\newcommand{\sigOmBIX      }{0.011}	 \newcommand{\degOmBIX      }{9\%}
\newcommand{\sigOmBX       }{0.010}	 \newcommand{\degOmBX       }{0\%}
\newcommand{\sigOmBXI      }{0.009}	 \newcommand{\degOmBXI      }{7\%}
\newcommand{\sigOmBXII     }{0.009}	 \newcommand{\degOmBXII     }{12\%}
\newcommand{\sigOmBXIII    }{0.008}	 \newcommand{\degOmBXIII    }{16\%}
\newcommand{\sigOmBXIV     }{0.008}	 \newcommand{\degOmBXIV     }{17\%}
\newcommand{\sigOmBXV      }{0.008}	 \newcommand{\degOmBXV      }{19\%}
\newcommand{\sigOmBXVI     }{0.008}	 \newcommand{\degOmBXVI     }{20\%}
\newcommand{\sigOmBXVII    }{0.008}	 \newcommand{\degOmBXVII    }{21\%}
\newcommand{\sigOmBXVIII   }{0.008}	 \newcommand{\degOmBXVIII   }{22\%}
\newcommand{\sigOmBXIX     }{0.008}	 \newcommand{\degOmBXIX     }{22\%}
\newcommand{\sigOmBXX      }{0.008}	 \newcommand{\degOmBXX      }{22\%}

\newcommand{\sigwBIII     }{0.201}	 \newcommand{\degwBIII     }{341\%}
\newcommand{\sigwBIV      }{0.125}	 \newcommand{\degwBIV      }{173\%}
\newcommand{\sigwBV       }{0.089}	 \newcommand{\degwBV       }{95\%}
\newcommand{\sigwBVI      }{0.073}	 \newcommand{\degwBVI      }{59\%}
\newcommand{\sigwBVII     }{0.068}	 \newcommand{\degwBVII     }{48\%}
\newcommand{\sigwBVIII    }{0.057}	 \newcommand{\degwBVIII    }{24\%}
\newcommand{\sigwBIX      }{0.050}	 \newcommand{\degwBIX      }{10\%}
\newcommand{\sigwBX       }{0.046}	 \newcommand{\degwBX       }{0\%}
\newcommand{\sigwBXI      }{0.042}	 \newcommand{\degwBXI      }{7\%}
\newcommand{\sigwBXII     }{0.040}	 \newcommand{\degwBXII     }{12\%}
\newcommand{\sigwBXIII    }{0.038}	 \newcommand{\degwBXIII    }{16\%}
\newcommand{\sigwBXIV     }{0.036}	 \newcommand{\degwBXIV     }{20\%}
\newcommand{\sigwBXV      }{0.035}	 \newcommand{\degwBXV      }{22\%}
\newcommand{\sigwBXVI     }{0.035}	 \newcommand{\degwBXVI     }{24\%}
\newcommand{\sigwBXVII    }{0.034}	 \newcommand{\degwBXVII    }{24\%}
\newcommand{\sigwBXVIII   }{0.034}	 \newcommand{\degwBXVIII   }{26\%}
\newcommand{\sigwBXIX     }{0.033}	 \newcommand{\degwBXIX     }{27\%}
\newcommand{\sigwBXX      }{0.033}	 \newcommand{\degwBXX      }{28\%}

\newcommand{\sigOmCIII     }{0.051}	 \newcommand{\degOmCIII     }{334\%}
\newcommand{\sigOmCIV      }{0.034}	 \newcommand{\degOmCIV      }{185\%}
\newcommand{\sigOmCV       }{0.025}	 \newcommand{\degOmCV       }{110\%}
\newcommand{\sigOmCVI      }{0.020}	 \newcommand{\degOmCVI      }{67\%}
\newcommand{\sigOmCVII     }{0.017}	 \newcommand{\degOmCVII     }{47\%}
\newcommand{\sigOmCVIII    }{0.015}	 \newcommand{\degOmCVIII    }{24\%}
\newcommand{\sigOmCIX      }{0.013}	 \newcommand{\degOmCIX      }{10\%}
\newcommand{\sigOmCX       }{0.012}	 \newcommand{\degOmCX       }{0\%}
\newcommand{\sigOmCXI      }{0.011}	 \newcommand{\degOmCXI      }{7\%}
\newcommand{\sigOmCXII     }{0.010}	 \newcommand{\degOmCXII     }{11\%}
\newcommand{\sigOmCXIII    }{0.010}	 \newcommand{\degOmCXIII    }{15\%}
\newcommand{\sigOmCXIV     }{0.010}	 \newcommand{\degOmCXIV     }{16\%}
\newcommand{\sigOmCXV      }{0.010}	 \newcommand{\degOmCXV      }{18\%}
\newcommand{\sigOmCXVI     }{0.009}	 \newcommand{\degOmCXVI     }{20\%}
\newcommand{\sigOmCXVII    }{0.009}	 \newcommand{\degOmCXVII    }{21\%}
\newcommand{\sigOmCXVIII   }{0.009}	 \newcommand{\degOmCXVIII   }{22\%}
\newcommand{\sigOmCXIX     }{0.009}	 \newcommand{\degOmCXIX     }{22\%}
\newcommand{\sigOmCXX      }{0.009}	 \newcommand{\degOmCXX      }{23\%}

\newcommand{\sigwCIII     }{0.254}	 \newcommand{\degwCIII     }{420\%}
\newcommand{\sigwCIV      }{0.166}	 \newcommand{\degwCIV      }{239\%}
\newcommand{\sigwCV       }{0.100}	 \newcommand{\degwCV       }{105\%}
\newcommand{\sigwCVI      }{0.084}	 \newcommand{\degwCVI      }{71\%}
\newcommand{\sigwCVII     }{0.077}	 \newcommand{\degwCVII     }{58\%}
\newcommand{\sigwCVIII    }{0.062}	 \newcommand{\degwCVIII    }{27\%}
\newcommand{\sigwCIX      }{0.054}	 \newcommand{\degwCIX      }{10\%}
\newcommand{\sigwCX       }{0.049}	 \newcommand{\degwCX       }{0\%}
\newcommand{\sigwCXI      }{0.046}	 \newcommand{\degwCXI      }{6\%}
\newcommand{\sigwCXII     }{0.044}	 \newcommand{\degwCXII     }{10\%}
\newcommand{\sigwCXIII    }{0.042}	 \newcommand{\degwCXIII    }{14\%}
\newcommand{\sigwCXIV     }{0.040}	 \newcommand{\degwCXIV     }{19\%}
\newcommand{\sigwCXV      }{0.039}	 \newcommand{\degwCXV      }{21\%}
\newcommand{\sigwCXVI     }{0.038}	 \newcommand{\degwCXVI     }{22\%}
\newcommand{\sigwCXVII    }{0.037}	 \newcommand{\degwCXVII    }{24\%}
\newcommand{\sigwCXVIII   }{0.036}	 \newcommand{\degwCXVIII   }{25\%}
\newcommand{\sigwCXIX     }{0.036}	 \newcommand{\degwCXIX     }{27\%}
\newcommand{\sigwCXX      }{0.035}	 \newcommand{\degwCXX      }{28\%}

\newcommand{\degcpOmAIII     }{224\%}	 \newcommand{\degcpwAIII     }{82\%}
\newcommand{\degcpOmAIV      }{133\%}	 \newcommand{\degcpwAIV      }{89\%}
\newcommand{\degcpOmAV       }{71\%}	 \newcommand{\degcpwAV       }{45\%}
\newcommand{\degcpOmAVI      }{47\%}	 \newcommand{\degcpwAVI      }{40\%}
\newcommand{\degcpOmAVII     }{40\%}	 \newcommand{\degcpwAVII     }{38\%}
\newcommand{\degcpOmAVIII    }{34\%}	 \newcommand{\degcpwAVIII    }{24\%}
\newcommand{\degcpOmAIX      }{30\%}	 \newcommand{\degcpwAIX      }{16\%}
\newcommand{\degcpOmAX       }{27\%}	 \newcommand{\degcpwAX       }{12\%}
\newcommand{\degcpOmAXI      }{25\%}	 \newcommand{\degcpwAXI      }{14\%}
\newcommand{\degcpOmAXII     }{22\%}	 \newcommand{\degcpwAXII     }{17\%}
\newcommand{\degcpOmAXIII    }{22\%}	 \newcommand{\degcpwAXIII    }{20\%}
\newcommand{\degcpOmAXIV     }{22\%}	 \newcommand{\degcpwAXIV     }{18\%}
\newcommand{\degcpOmAXV      }{23\%}	 \newcommand{\degcpwAXV      }{20\%}
\newcommand{\degcpOmAXVI     }{23\%}	 \newcommand{\degcpwAXVI     }{23\%}
\newcommand{\degcpOmAXVII    }{22\%}	 \newcommand{\degcpwAXVII    }{26\%}
\newcommand{\degcpOmAXVIII   }{23\%}	 \newcommand{\degcpwAXVIII   }{26\%}
\newcommand{\degcpOmAXIX     }{23\%}	 \newcommand{\degcpwAXIX     }{24\%}
\newcommand{\degcpOmAXX      }{24\%}	 \newcommand{\degcpwAXX      }{23\%}

\newcommand{\degcpOmBIII     }{35\%}	 \newcommand{\degcpwBIII     }{69\%}
\newcommand{\degcpOmBIV      }{23\%}	 \newcommand{\degcpwBIV      }{46\%}
\newcommand{\degcpOmBV       }{15\%}	 \newcommand{\degcpwBV       }{28\%}
\newcommand{\degcpOmBVI      }{12\%}	 \newcommand{\degcpwBVI      }{19\%}
\newcommand{\degcpOmBVII     }{11\%}	 \newcommand{\degcpwBVII     }{14\%}
\newcommand{\degcpOmBVIII    }{8\%}	 \newcommand{\degcpwBVIII    }{7\%}
\newcommand{\degcpOmBIX      }{5\%}	 \newcommand{\degcpwBIX      }{3\%}
\newcommand{\degcpOmBX       }{4\%}	 \newcommand{\degcpwBX       }{2\%}
\newcommand{\degcpOmBXI      }{3\%}	 \newcommand{\degcpwBXI      }{4\%}
\newcommand{\degcpOmBXII     }{3\%}	 \newcommand{\degcpwBXII     }{7\%}
\newcommand{\degcpOmBXIII    }{4\%}	 \newcommand{\degcpwBXIII    }{12\%}
\newcommand{\degcpOmBXIV     }{6\%}	 \newcommand{\degcpwBXIV     }{15\%}
\newcommand{\degcpOmBXV      }{8\%}	 \newcommand{\degcpwBXV      }{22\%}
\newcommand{\degcpOmBXVI     }{11\%}	 \newcommand{\degcpwBXVI     }{28\%}
\newcommand{\degcpOmBXVII    }{14\%}	 \newcommand{\degcpwBXVII    }{35\%}
\newcommand{\degcpOmBXVIII   }{16\%}	 \newcommand{\degcpwBXVIII   }{39\%}
\newcommand{\degcpOmBXIX     }{18\%}	 \newcommand{\degcpwBXIX     }{43\%}
\newcommand{\degcpOmBXX      }{19\%}	 \newcommand{\degcpwBXX      }{45\%}

\newcommand{\degcpOmCIII     }{79\%}	 \newcommand{\degcpwCIII     }{77\%}
\newcommand{\degcpOmCIV      }{54\%}	 \newcommand{\degcpwCIV      }{71\%}
\newcommand{\degcpOmCV       }{38\%}	 \newcommand{\degcpwCV       }{33\%}
\newcommand{\degcpOmCVI      }{29\%}	 \newcommand{\degcpwCVI      }{31\%}
\newcommand{\degcpOmCVII     }{25\%}	 \newcommand{\degcpwCVII     }{28\%}
\newcommand{\degcpOmCVIII    }{20\%}	 \newcommand{\degcpwCVIII    }{14\%}
\newcommand{\degcpOmCIX      }{17\%}	 \newcommand{\degcpwCIX      }{6\%}
\newcommand{\degcpOmCX       }{15\%}	 \newcommand{\degcpwCX       }{3\%}
\newcommand{\degcpOmCXI      }{14\%}	 \newcommand{\degcpwCXI      }{3\%}
\newcommand{\degcpOmCXII     }{13\%}	 \newcommand{\degcpwCXII     }{4\%}
\newcommand{\degcpOmCXIII    }{13\%}	 \newcommand{\degcpwCXIII    }{6\%}
\newcommand{\degcpOmCXIV     }{14\%}	 \newcommand{\degcpwCXIV     }{7\%}
\newcommand{\degcpOmCXV      }{15\%}	 \newcommand{\degcpwCXV      }{10\%}
\newcommand{\degcpOmCXVI     }{17\%}	 \newcommand{\degcpwCXVI     }{13\%}
\newcommand{\degcpOmCXVII    }{18\%}	 \newcommand{\degcpwCXVII    }{15\%}
\newcommand{\degcpOmCXVIII   }{19\%}	 \newcommand{\degcpwCXVIII   }{16\%}
\newcommand{\degcpOmCXIX     }{19\%}	 \newcommand{\degcpwCXIX     }{16\%}
\newcommand{\degcpOmCXX      }{20\%}	 \newcommand{\degcpwCXX      }{16\%}


\newcommand{\sigIII     }{(\sigOmBIII     , \sigwBIII     )}
\newcommand{\sigIV      }{(\sigOmBIV      , \sigwBIV      )}
\newcommand{\sigV       }{(\sigOmBV       , \sigwBV       )}
\newcommand{\sigVI      }{(\sigOmBVI      , \sigwBVI      )}
\newcommand{\sigVII     }{(\sigOmBVII     , \sigwBVII     )}
\newcommand{\sigVIII    }{(\sigOmBVIII    , \sigwBVIII    )}
\newcommand{\sigIX      }{(\sigOmBIX      , \sigwBIX      )}
\newcommand{\sigX       }{(\sigOmBX       , \sigwBX       )}
\newcommand{\sigXI      }{(\sigOmBXI      , \sigwBXI      )}
\newcommand{\sigXII     }{(\sigOmBXII     , \sigwBXII     )}
\newcommand{\sigXIII    }{(\sigOmBXIII    , \sigwBXIII    )}
\newcommand{\sigXIV     }{(\sigOmBXIV     , \sigwBXIV     )}
\newcommand{\sigXV      }{(\sigOmBXV      , \sigwBXV      )}
\newcommand{\sigXVI     }{(\sigOmBXVI     , \sigwBXVI     )}
\newcommand{\sigXVII    }{(\sigOmBXVII    , \sigwBXVII    )}
\newcommand{\sigXVIII   }{(\sigOmBXVIII   , \sigwBXVIII   )}
\newcommand{\sigXIX     }{(\sigOmBXIX     , \sigwBXIX     )}
\newcommand{\sigXX      }{(\sigOmBXX      , \sigwBXX      )}

\newcommand{\degIII     }{(\degOmBIII     , \degwBIII     )}
\newcommand{\degIV      }{(\degOmBIV      , \degwBIV      )}
\newcommand{\degV       }{(\degOmBV       , \degwBV       )}
\newcommand{\degVI      }{(\degOmBVI      , \degwBVI      )}
\newcommand{\degVII     }{(\degOmBVII     , \degwBVII     )}
\newcommand{\degVIII    }{(\degOmBVIII    , \degwBVIII    )}
\newcommand{\degIX      }{(\degOmBIX      , \degwBIX      )}
\newcommand{\degX       }{(\degOmBX       , \degwBX       )}
\newcommand{\degXI      }{(\degOmBXI      , \degwBXI      )}
\newcommand{\degXII     }{(\degOmBXII     , \degwBXII     )}
\newcommand{\degXIII    }{(\degOmBXIII    , \degwBXIII    )}
\newcommand{\degXIV     }{(\degOmBXIV     , \degwBXIV     )}
\newcommand{\degXV      }{(\degOmBXV      , \degwBXV      )}
\newcommand{\degXVI     }{(\degOmBXVI     , \degwBXVI     )}
\newcommand{\degXVII    }{(\degOmBXVII    , \degwBXVII    )}
\newcommand{\degXVIII   }{(\degOmBXVIII   , \degwBXVIII   )}
\newcommand{\degXIX     }{(\degOmBXIX     , \degwBXIX     )}
\newcommand{\degXX      }{(\degOmBXX      , \degwBXX      )}

\newcommand{\degcpOmIII     }{\degcpOmBIII     }
\newcommand{\degcpOmIV      }{\degcpOmBIV      }
\newcommand{\degcpOmV       }{\degcpOmBV       }
\newcommand{\degcpOmVI      }{\degcpOmBVI      }
\newcommand{\degcpOmVII     }{\degcpOmBVII     }
\newcommand{\degcpOmVIII    }{\degcpOmBVIII    }
\newcommand{\degcpOmIX      }{\degcpOmBIX      }
\newcommand{\degcpOmX       }{\degcpOmBX       }
\newcommand{\degcpOmXI      }{\degcpOmBXI      }
\newcommand{\degcpOmXII     }{\degcpOmBXII     }
\newcommand{\degcpOmXIII    }{\degcpOmBXIII    }
\newcommand{\degcpOmXIV     }{\degcpOmBXIV     }
\newcommand{\degcpOmXV      }{\degcpOmBXV      }
\newcommand{\degcpOmXVI     }{\degcpOmBXVI     }
\newcommand{\degcpOmXVII    }{\degcpOmBXVII    }
\newcommand{\degcpOmXVIII   }{\degcpOmBXVIII   }
\newcommand{\degcpOmXIX     }{\degcpOmBXIX     }
\newcommand{\degcpOmXX      }{\degcpOmBXX      }

\newcommand{\degcpwIII     }{\degcpwBIII     }
\newcommand{\degcpwIV      }{\degcpwBIV      }
\newcommand{\degcpwV       }{\degcpwBV       }
\newcommand{\degcpwVI      }{\degcpwBVI      }
\newcommand{\degcpwVII     }{\degcpwBVII     }
\newcommand{\degcpwVIII    }{\degcpwBVIII    }
\newcommand{\degcpwIX      }{\degcpwBIX      }
\newcommand{\degcpwX       }{\degcpwBX       }
\newcommand{\degcpwXI      }{\degcpwBXI      }
\newcommand{\degcpwXII     }{\degcpwBXII     }
\newcommand{\degcpwXIII    }{\degcpwBXIII    }
\newcommand{\degcpwXIV     }{\degcpwBXIV     }
\newcommand{\degcpwXV      }{\degcpwBXV      }
\newcommand{\degcpwXVI     }{\degcpwBXVI     }
\newcommand{\degcpwXVII    }{\degcpwBXVII    }
\newcommand{\degcpwXVIII   }{\degcpwBXVIII   }
\newcommand{\degcpwXIX     }{\degcpwBXIX     }
\newcommand{\degcpwXX      }{\degcpwBXX      }

\title{Effects of Completeness and Purity on Cluster Dark Energy Constraints}
\author{Michel Aguena}
\email{aguena@if.usp.br}
\author{Marcos Lima} 
\affiliation{
Departamento de F\'{\i}sica Matem\'atica, 
Instituto de F\'{\i}sica, Universidade de S\~ao Paulo, 
CP 66318, CEP 05314-970, S\~ao Paulo, SP, Brazil 
}
%
%

\date{\today}

\begin{abstract}
\baselineskip 11pt
The statistical properties of galaxy clusters can only be used for cosmological purposes 
if observational effects related to cluster detection are accurately characterized. 
These effects include the 
selection function \fixed{associated with} cluster finder algorithms and survey strategy. 
The importance of the selection becomes apparent when different 
cluster finders are applied to the same galaxy catalog, producing different cluster 
samples. 
We consider parametrized functional forms for the observable-mass relation, 
its scatter as well as the completeness and purity of cluster samples, and study how prior knowledge on these 
function parameters affects dark energy constraints derived from cluster statistics.  
Under the assumption 
\fixedII{of a fiducial model for the selection function where}
completeness and purity reach 
50\% at masses around $10^{13.5}M_{\odot}/h$, we find that self-calibration of selection parameters in current and upcoming cluster surveys is possible, while still allowing for competitive dark energy constraints. We consider a fiducial survey with specifications similar to those of the Dark Energy Survey (DES) with 5000 deg$^2$, maximum redshift of $z_{max}\sim 1.0$ and threshold observed mass $M_{th}\sim 10^{13.8}M_{\odot}/h$, such that completeness and purity $\sim 60\%-80\%$ at masses around $M_{th}$. 
Perfect knowledge of all selection parameters allows for constraining a constant dark energy equation of state 
to $\sigma(w)=\sigxxwB$. Employing a joint fit including self-calibration of 
the effective selection degrades 
constraints to $\sigma(w)=\sigffwB$. External calibrations at the level of \prv\% in the 
parameters of the observable-mass relation and completeness/purity functions are necessary to 
improve the joint constraints to $\sigma(w)=\sigppwB$.
In the lack of knowledge of selection parameters, future experiments probing larger areas and greater depths suffer from stronger relative degradations on dark energy constraints compared to
current surveys.  

\end{abstract}

\maketitle

\section{Introduction} \label{sec:intro}

The properties of dark matter halos have been characterized with increasing accuracy  
through dark matter N-body simulations 
of multiple cosmological models 
\cite{MilleniumII09, Kly11, Kly16, Fos15, Cro15,Fos15.2,Schmidt09}. 
\fixed{However, clusters of galaxies,}
observed in surveys spanning different wavelengths carry a number 
of observational effects 
\cite{LimHu04, LimHu05, LimHu07,Wu08, Eri11,Rozo11, Asc16, Asc16b}. 
For the cosmological use of galaxy clusters,
\fixed{it is necessary to understand these effects in detail (e.g. by measuring them in simulations) and use this knowledge to parametrize the effects as appropriate functions of intrinsic cluster parameters (e.g. mass and redshift).}
An ideal self-consistent analysis must 
then constrain both cosmological parameters of interest as well as {\it nuisance} parameters 
related to astrophysical and observational effects, despite intrinsic degeneracies 
\cite{LevSchWhi02, MajMoh04, LimHu04, LimHu05, LimHu07, Gladders07, Cunha08, Cunha09, Rozo10, 
Benson13, Planck14}. 
In this context, external calibrations of {\it nuisance} 
parameters may help improve constraining cosmology.  

Given a set of true halos and the matter tracers associated to them (e.g. optical galaxies), 
the first step is to characterize the performance of algorithms for cluster identification
\fixed{
called cluster finders.
Some of these methods,
such as \code{MaxBCG}\cite{MaxBCG}, \code{FoF} used in \cite{FarrensAbdallaCyprianoCF} and \code{RedMaPPer}\cite{RedMapper14, Rykoff16},
are based on the presence of red-sequence galaxies within clusters.
They have the advantage of including this extra information, which is certainly valuable at low redshifts. However they may suffer from limitations at higher redshifts.
Meanwhile, there are cluster finders such as \code{WAZP}\cite{WAZP}, \code{VT}\cite{VT11}
and \code{C4}\cite{C4},
which rely mostly on detecting spatial overdensities.
These algorithms thrive to provide better detections at higher redshifts,
although they typically depend more strongly on the quality of galaxy photometric redshifts.
}
Cluster finders may fail to identify a fraction of clusters related to dark matter halos,
as well as detect false clusters with no association to halos.  
These two problems can be quantified by the so-called {\it completeness} and {\it purity} 
of the cluster sample 
\cite{VT11, Asc16, Asc16b}, which typically reflect 
limitations of the cluster finder algorithm, such as e.g. artificial over-merging or fragmentation of 
clusters relative to their corresponding halos.
Whereas completeness and purity may depend on various factors -- such as 
survey specifications, the quality of photometric 
redshifts (photo-zs) and the observable-mass relation -- they 
are mainly properties of the cluster finder itself. 
We will often refer to completeness and purity as describing the cluster selection function. 

Next we must  consider the {\it observable-mass} relation, typically 
characterized by a mean relation and a scatter 
\cite{LevSchWhi02, MajMoh04, LimHu04, LimHu05}. 
For optical clusters the observable is 
the \fixed{cluster richness, representing} the number of cluster member galaxies. 
Richness may also refer to a subsample of member galaxies whose properties more closely relate to halo mass 
(e.g. richness can be based on red-sequence galaxies within a cluster \cite{RedMapper14, Rykoff16}, as opposed to all member galaxies ). 
In simulations, 
clusters correctly matched to dark matter halos can be used to characterize the observable-mass relation
\cite{Nag06,Kra06,Nag07,Old15,Yu15,Asc16}. 
\fixed{However, 
it is important to point out that all of these calibrations are not enough to determine observable-mass and selection parameters at the percent level, 
even though they are still useful to help determine at least appropriate function forms or loose priors for parameters, 
which are then self-calibrated in a full cosmological analysis.}
Observationally, optical clusters may be matched to 
detections in other wavelengths (e.g. millimeter or X-ray) from which observable-observable scaling 
relations can be estimated \cite{Bon08, Rykoff08, Sar15}, 
and under the assumption of hydrostatic 
equilibrium, 
observable-mass relations may be derived.
Alternatively, lensing masses may be 
available for a fraction of the optical clusters 
\cite{Becker11, Gruen14, vonderLinden14, vonderLinden14b, Die14, Battaglia16, PennaLima16}. 
In conjunction, simulations and observational cross-matches allow for independent external calibrations of the 
observable-mass relation. 

The scatter in the observable-mass relation 
may also be assessed from simulations and observations, and it can be tied to different sources 
\cite{Rozo11}. 
An intrinsic scatter exists even for a perfect cluster finder (i.e. one with unit 
completeness and purity) and represents instances where a cluster of given richness 
has a range of masses due to intrinsic variability in the physical processes that 
relate these quantities, turning them stochastic
\cite{Nag06,Kra06}.
On the other hand, imperfections in the matching of clusters may artificially change this otherwise intrinsic scatter, as well as 
other observational issues \cite{Rozo11}. 
We will also refer to the {\it effective} selection function, which is characterized by a 
combination of the actual 
selection function (completeness/purity) and the observable-mass relation.  
 
There may be an interplay between the derived observable-mass relation 
and the sample selection function, as the characterization of both depend on the 
matching process of clusters and halos (in simulations) and,
clusters and clusters (in multi-wavelength observations). 
For instance, clusters catastrophically scattered in and out of a given richness bin may affect the sample 
completeness and purity, an effect which may be parametrized by altering the observable-mass 
distribution to include an extra Gaussian term 
\cite{Eri11}. Conversely, using only clusters and/or halos which are believed to have been correctly 
matched to define the observable-mass relation may produce a relation with unrealistically low scatter. 
Despite these issues, it is conceptually simpler to keep the definitions of completeness and 
purity decoupled from the observable-mass relation, 
and we will follow this approach in this work by parameterizing these functions independently
\fixed{
using functional forms characterized in simulations \cite{Aguena_etal_inprep}.
}.

Finally, we must characterize errors in the cluster photometric redshifts (photo-zs) 
\cite{LimHu07}. 
We will again take the simpler approach of decoupling photo-z errors from completeness and purity issues, as photo-z errors are mainly tied 
to degeneracies in color-magnitude-redshift space and the efficiency of photo-z algorithms
\cite{Oya08b, Abd11, San14}.
The selection function of cluster finders that make direct use of photo-zs
\cite{VT11,Die14} is clearly 
affected by the photo-z quality, which may translate to additional sources of over-merging and fragmentation of 
clusters in the line-of-sight. However, for cluster 
galaxies we expect the photo-z errors to be considerably smaller than for field galaxies. 
Therefore in this work we will neglect such effects, 
as our goal is to assess the direct impact of 
completeness and purity issues on cluster cosmology.
Our analysis is conservative in this sense, since by 
including the extra dependencies of completeness 
and purity on observable-mass and photo-z parameters 
would effectively decrease the number of {\it nuisance} parameters to constrain, 
potentially increasing the sensitivity of cluster observables. 

In this paper we study how the inclusion of the cluster sample completeness and purity 
impacts the cosmological constraints derived from that sample. 
For a given parametrization of these functions, we 
also explore how prior knowledge on the selection 
can help constrain dark energy parameters in current and upcoming galaxy 
surveys. 
We start in \S~\ref{sec:completeness} discussing the characterization of the 
selection function via the sample completeness and purity. 
In \S~\ref{sec:counts} we discuss the formalism for predicting 
cluster counts and covariance, including selection effects. 
In \S~\ref{sec:fisher} we detail the Fisher Matrix formalism 
to predict dark energy constraints and biases from cluster statistics and in \S~\ref{sec:fiducial}
we present the fiducial model, including selection parametrizations. In \S~\ref{sec:results} we present our main results 
and in \S~\ref{sec:discussion} we discuss these results and conclude.

\section{Completeness and Purity} \label{sec:completeness}

We define the {\it completeness} of a cluster catalog as the fraction of galaxy 
clusters {\it correctly} identified relative to the number of {\it true} dark matter halos. 
Likewise, the {\it purity}  of the same catalog is defined as the fraction of galaxy clusters 
correctly identified relative to the {\it total} number of detected clusters. 
Clearly both concepts are important to characterize the cluster 
finder selection function, since nearly all algorithms lead to 
samples that are both incomplete and impure in certain 
ranges of masses and redshifts. 
A low completeness indicates an inefficiency of the cluster finder in detecting 
systems that it should have detected (or which a perfect cluster finder detects), 
whereas a low purity indicates a high fraction of false-positives in the sample, i.e. 
detections incorrectly made (and which a perfect cluster 
finder would not have made).
  
The completeness and purity of a cluster finder depend on the assumptions 
it makes and also on observing conditions of a specific survey. 
For instance, a cluster finder which uses information from 
the galaxy red-sequence -- observed in most low-redshift clusters -- 
has the possibility of outperforming a cluster finder that ignores 
this information. On the other hand, 
if the assumption of a red-sequence is extrapolated into a domain 
in which it may not apply (e.g. at higher redshifts), such cluster 
finder may produce samples that are either incomplete or impure. 
As a result, different performances may be observed when comparing 
different cluster finders applied to the same data 
set as well as the same cluster finder applied to different surveys. 

From the above definitions of completeness and purity, these quantities 
require matching clusters to dark matter halos. Strictly speaking 
this can only be directly assessed in simulated catalogs, where 
information about the true underlying dark matter halos is fully available. 
However cross-checks from real observations may 
also provide useful hints into the selection function of a given 
cluster finder.  
Here we will assume that 
simulations representative of the observing conditions 
are available for purposes of roughly estimating the cluster finder 
selection function as well as the observable-mass relation. 
Clearly, simulations of this kind necessarily make certain assumptions 
that may not apply to real observed data. Nonetheless they are 
useful to roughly calibrate cluster finders and estimates of 
their selection under these assumptions. 
When performing a cosmological analysis on real data, one would not 
fully trust simulation results, but they 
might inspire functional forms for parameterizing the 
cluster selection and observable-mass relation \cite{Kra06,Nag06}, whose 
parameter values can then be obtained from a self-consistent cosmological 
analysis of the cluster sample.

For pedagogical reasons, let us outline the process of using a  
simulated galaxy catalog and its associated dark matter halos 
for defining the cluster sample completeness, 
purity and observable-mass relation. 
Given the list of true dark matter halos of mass $M$ and  redshift $z$, 
and the catalog of galaxies populating these halos, one may run a cluster finder 
producing a list of clusters with certain observed properties (e.g. richness 
and photo-zs for optical clusters). Since the considerations made here apply 
to detections not only of optical clusters, but for multiple wavelengths, we will often refer to the 
{\it observed mass} $\Mobs$ instead of the direct observable ${\cal O}$.
\fixed{Our fiducial survey will be similar to the Dark Energy Survey (DES), so we will typically refer to the richness as the observable,
derived from optical cluster finders. 
However, all results also apply to cluster finders defined at other wavelengths with different mass proxies.
Here $\Mobs$ is the mass inferred from the observable,} 
being therefore equivalent to it, but in mass units. 
We will formally characterize individual clusters by their 
values of $\Mobs$ and photo-z, denoted $\zphot$, 
\fixed{ and this notation applies even when we consider an specific observable such as richness.}
Given the number $N_{\rm h}$ of halos found and the number $N_{\rm c}$ of 
clusters detected, we may then consider 
the following steps towards characterizing the cluster finder selection function and the observable-mass distribution:

\begin{itemize} 
\item Rank the $N_{\rm h}$ halos by mass $M$ and the $N_{\rm c}$ clusters by richness ${\cal O}$.
\item Perform a matching of halos and clusters, producing $N_{\rm mat}$ matches.
\fixedII{
\begin{itemize} 
\item In general the matching of halos and clusters can be done in two ways: by the fraction of coincident halo and cluster members, or by the spatial proximity of halo and cluster centers (either 3D or angular). 
\item In cases where multiple matches may potentially occur (e.g. within a cluster radius one finds more than one halo center), the most massive halo or richest cluster may be selected among the possible candidates.
This assures that e.g. the most massive halo that is spatially close to the richest unmatched cluster will preferentially produce a match.
\item A two-way matching can be applied to reassure a more stringent matching criterion.
In this case the matching is made in both directions (halos are matched to clusters and vice-versa) and only pairs that coincide in both directions are kept as true matches. 
\end{itemize}
}  
\item Plot ${\cal O}$ versus $M$ for the matches to determine 
the observable-mass relation and its scatter. A cluster mass computed from 
this relation using the value of the observable ${\cal O}$ represents the cluster observed mass $\Mobs$.
\item For each bin of halo mass $M$ and redshift $z$, compute the completeness $c(M, z)$ as
\bea
c(M, z)=\frac{N_{\rm mat}(M, z)}{N_{\rm h}(M, z)}\,.
\eea
\item For each bin of cluster observed mass $\Mobs$ and photo-z $\zphot$, compute the 
purity $p(\Mobs, \zphot)$ as
\bea
p(\Mobs, \zphot)=\frac{N_{\rm mat}(\rich, \zphot)}{N_{\rm c}(\rich, \zphot)}\,.
\eea
\end{itemize}

Clearly these definitions depend on the \fixed{specific matching criterion} 
imposed in the second step above (see further discussion on a related issue in \S~\ref{subsec:clustercounts}). {\bf Notice that 
the ranking of halos and clusters in the very first step plays only a secondary role in the matching and is in fact dispensable. It enters only as an additional criterium to resolve potential multiple matches according to physical matching criteria defined in the second step, either by membership overlap or spatial proximity.}

We may also use these matches to estimate cluster $\zphot$ errors, 
which depend both on the quality of galaxy photo-zs 
and on the cluster finder performance in assigning 
redshifts to clusters. 
In this work, we will assume that the effect of photo-z errors is already encapsulated 
into the estimated completeness and purity and does not represent 
an extra source of cosmological degeneracies \cite{LimHu07}. 
Obviously such assumption should be checked for each cluster finder, especially for those which heavily rely on galaxy photo-z estimates.

\fixed{
In observed data,
the estimation of completeness and purity becomes intrinsically more complicated,
as the mass of the clusters is not known and 
because no observed catalog can be taken as a truth table.
Although the mass of the clusters can be estimated via observable-mass relations,
the lack of a truth table makes the extraction of completeness and purity information from the data alone currently inviable.
If reliable mock catalogs for a given survey are not available, 
calibration of scaling relations is possible from lensing masses 
measured for a fraction of the detected clusters or 
from matching e.g. optical clusters to detections at other wavelengths. 
Thus, 
we can obtain limited information about the observable-mass 
relation and its scatter. 
In the worst-case scenario, we could assume 
a generic selection function and fully self-calibrate its parameters 
(that is, to constrain the parameters along with the cosmology) 
from the observed cluster data alone. 
Fortunately, we expect reliable simulations, lensing masses, 
multiple external cross-calibrations and spectroscopic follow-ups 
to be available for a self-consistent 
cosmological analysis of most current and future cluster surveys.  
}

\section{Observed Cluster Properties} \label{sec:counts}

\subsection{Cluster Counts} \label{subsec:clustercounts}

We parametrize the theoretical dark matter halo mass-function as 
\eqn{
\frac{d \bar n (z,M)}{ d\ln M}  = \frac{\bar{\rho}_m}{M} \ \frac{d\ln{\sigma^{-1}}}{d\ln{M}} \ f(\sigma) \,,
\label{eq:massfunc}
}
\noindent
where $\sigma^2(M,z)$ is the variance of the linear density field in a 
spherical region of radius $R$ enclosing a mass $M=4\pi R^3\bar{\rho}_m/3$ at the present background matter density $\bar{\rho}_m$. We take $f(\sigma)$ from a fit to simulations by Tinker et al \cite{Tin08}, 
with parameter values appropriate for overdensity $\Delta=200$ with respect to the background matter density.
The predicted comoving number density $\bar{n}_{\alpha}$ of clusters in the \fixed{observed-mass bin (indexed by $\alpha$)} is obtained by integrating the 
mass function convolved with all observational effects mentioned previously
as \cite{LimHu04, LimHu05,LimHu07}
\bea
\bar{n}_{\alpha}(z)&=& \int_{\rich_{\alpha}}^{\rich_{\alpha+1}}  d\ln \rich    \int_{0}^{\infty}  d\ln M \frac{d\bar{n}_{\rm obs}}{d\ln M}\,,
\label{eq:numdens}
\eea
where the {\it observed} mass-function
\bea
\frac{d\bar{n}_{\rm obs}}{d\ln M}=
\frac{d\bar{n}(z,M)}{d\ln M} 
P(\rich|M) \frac{c(M,z)}{p(\rich,\zphot)}\,
\eea
carries the effects of completeness, purity and observable-mass distribution $P(\Mobs |M)$, assumed to be Gaussian in $\ln M$.
The number counts in the $i^{\rm th}$ photo-z bin are then obtained by 
integrating the comoving number density in comoving volume or redshift, including 
the photo-z error distribution as \cite{LimHu07}
\bea
\bar{m}_{\alpha, i}&=& \int_{\zphot_i}^{\zphot_{i+1}} d\zphot  \int_{0}^{\infty} dz P(\zphot | z ) \frac{r^2(z)}{H^2(z)}
\bar{n}_{\alpha}(z) \,, 
\label{eq:numcounts}
\eea
where $H(z)$ is the 
Hubble parameter at redshift $z$ and $r(z)$ is the comoving angular 
diameter distance, identified here with the comoving radial 
distance since we only consider flat cosmologies. 
As mentioned previously, we will not consider the effect of photo-z errors 
explicitly here. In the above description, this implies taking 
$P(\zphot|z)$ to be a Dirac delta function, which then allows us to 
perform one of the redshift integrals trivially. In this case, we denote the {\it effective} cluster selection $f(\Mobs|M)$ as the combination
\fixed{
\bea
f(\rich,z|M)=P(\rich|M)\frac{c(M,z)}{p(\rich,z)}\,.
\label{eq:cef}
\eea
}

The separation of $f$ into these three components is mostly pedagogical, as the 
effective selection itself can be measured directly from simulations (with no reference to separate components). 
In fact, 
it is possible to consider completeness and purity effects (partially) as a propagation of projection effects into the otherwise intrinsic 
observable-mass relation $P(\rich|M)$, 
turning it into $f(\rich|M)$ \cite{Eri11}. 
Whereas simulations indicate that $P(\rich | M)$ can be parametrized as a log-gaussian distribution \cite{Nag06,Nag07,Kra06,Sar15,PennaLima16,Baxter16} with observable-mass 
relations displaying low scatter \cite{Kra06}, 
$f(\rich|M)$ would then have an extra 
log-gaussian component, turning the final distribution to be non-log-gaussian \cite{Eri11}. 

\fixed{Projection effects occur mainly as a result of photometric redshift errors,  
which cause cluster finders to fail in multiple ways. The simplest failure mode is when the cluster finder still detects individual clusters appropriately, but either includes field galaxies as cluster members or excludes true cluster members. In this case, projection effects conserve the total number of clusters and act merely as an extra source of scatter for richness estimates and could indeed be modeled as an extra log-gaussian component for the observable-mass distribution. However, in more extreme cases of failure, the cluster finder may inappropriately fragment one cluster into two, merge two separate clusters along the line of sight into a single cluster, or simply fail to find a cluster due to poor signal-to-noise. These failure modes do not conserve the number of clusters and more strongly affect the measured cluster counts and variance. Our parametrization of selection in terms of completeness and purity attempts to capture all these possible cluster-finder failure modes.}

\fixed{
As mentioned before,
the observable-mass relation $P(\rich|M)$ has an intrinsic scatter due the physical processes that correlate these quantities.
This scatter can be studied and quantified in simulations,
being well understood.
However,
there is an additional source of scatter related to the use of observed mass and richness on $P(\rich|M)$.
When calibrating $P(\rich|M)$,
the uncertainties on the measurement of both the richness and the mass also have to be included \cite{Sar15}.
For simplicity,
we will not consider this extra observational effect.
}

However, contamination by projection effects is not the only 
issue that may affect the total selection. One simple effect, which however 
is likely always present, is a mismatch between the effective overdensity $\Delta_c$ used to define observed clusters 
and the overdensity $\Delta_h$ of the dark matter halos 
associated to them (either halos directly matched to clusters in simulations 
or halos whose mass-function is used to predict the cluster abundance). 
For instance, if we use a halo mass-function appropriate for $\Delta_h$ to predict the {\it cluster} abundance as described 
above, but our cluster finder detects clusters at an effective overdensity 
$\Delta_c\ne \Delta_h$, a mismatch of halo and cluster properties will follow 
if not accounted for explicitly.
Notice that these effects 
may happen even for a perfect cluster finder, and because they are associated 
to the cluster detection itself, they cannot be corrected after detection by 
simply redefining cluster masses with a more appropriate overdensity or 
even a new observable. For clusters detected using signal-to-noise ratios 
or fixed apertures, which do not correspond to a fixed halo overdensity, 
it may be even trickier to interpret comparisons of cluster and halo properties.

From the considerations above, it is clear that completeness and purity depend on specific assumptions underlying cluster finder algorithms.
In this work we will parametrize the selection via separate functions for the sample completeness and purity as described in \S~\ref{sec:fiducial}.

\subsection{Cluster Covariance} \label{subsec:clustercov}

The local number counts $m_{\alpha, i}({\bf x})$ of clusters at position ${\bf x}$ fluctuate spatially around the mean predicted values $\bar{m}_{\alpha, i}$, following 
the matter density contrast $\delta({\bf x})$ as 
\bea
m_{\alpha,i}({\bf x})=
\bar{m}_{\alpha,i}
[1+ b_\alpha(z)\delta({\bf x})] \,,
\eea

\noindent
where $b_\alpha(z)$ is the average cluster bias defined as
\bea
b_\alpha(z)&=& \frac{1}{\bar{n}_\alpha(z)}
\int_0^\infty d\ln M \frac{d\bar{n}_\alpha}{d\ln M} b(M,z) \,. \nonumber \\
\eea

Notice that $b_\alpha(z)$ is consistently predicted from the number 
density in Eq.~\ref{eq:numdens}, and therefore carries the observable-mass 
and selection effects. 
Here $b(M,z)$ is the halo bias for which we will take a
fit to simulations by Tinker et al. 2010 \cite{Tin10} as
\eqn{
	b \de{M,z}=1-A\frac{\nu^a}{\nu^a+\delta_c^a}+B\nu^b+C\nu^c\,,
}
where $\nu(M,z)=\delta_c/\sigma(M,z)$, $\delta_c=1.686$
and we fix values for the parameters $A,B,C,a,b,c$ appropriate 
for the same overdensity $\Delta=200$ used in the abundance predictions.

The cluster counts have a sample covariance $S^{\al\be}_{ij}$ due to the 
large scale structure of the Universe given by \cite{HuKra2003,LimHu05}
\bea
S^{\al\be}_{ij}
&=&
\left<
\de{m_{\al,i}-\bar m_{\al,i}}
\de{m_{\be,j}-\bar m_{\be,j}}
\right>
\nonumber \\
&=&\bar m_{\al,i}b_{\al,i}\bar m_{\be,j}b_{\be,j}
\int \frac{d^3k}{\de{2\pi}^3}
P\de{k}
W_i^*({\bf k})
W_j({\bf k})\,, \nonumber \\
\label{eq:samp_cov}
\eea
where $W_i({\bf k})$ is the Fourier transform of the volume window 
function in bin $i$ and we set $b_{\alpha, i}\approx b_\alpha(z_i)$ 
at the bin centroid $z_i$, which is valid for sufficiently small 
redshift bins.

\newcommand{\kapar}{{k_\parallel}}
\newcommand{\kaper}{{k_\bot}}
\newcommand{\xapar}{{x_\parallel}}
\newcommand{\xaper}{{x_\bot}}

Here we will take a window to be a cylinder with a small angular radius ($\theta_s \lesssim 10$ deg) and height $\delta r_i$,
in which case $W_i({\bf k})$ is given by \cite{HuKra03, LimHu07}
\newcommand{\jz}[1]{j_0\de{#1}}
\eqn{
W_i({\bf k})=
2 \exp\left(i \kapar r_i\right)  
\jz{\frac{\kapar\delta r_i}{2}}  
\frac{J_1\de{\kaper \theta_s r_i}}{\kaper \theta_s r_i}
\,,}
where ${\bf k}=(k_\parallel, {\bf k}_\perp)$. 
The counts are also subject to Poisson variance or shot
noise given by
\eqn{
M^{\al\be}_{ij}
=\delta_{\al\be}\delta_{ij}\bar{m}_{\al,i}\,,
}
such that the total covariance $C^{\al\be}_{ij}$ is the sum of sample 
covariance and Poisson variance
\bea
C^{\al\be}_{ij}=S^{\al\be}_{ij}+M^{\al\be}_{ij}\,.
\eea

\section{Fisher Matrix} \label{sec:fisher}

We use the Fisher matrix formalism to study the effects of 
parametrizing the cluster selection function, given the predictions 
of cluster counts and covariance described in the previous section.
We split the counts in redshift, 
mass and angular cells. For convenience of notation, we let the index $i$ denote 
binning in photo-z, observed mass and angular pixel, and arrange 
the counts into a single vector ${\MM}$.  Similarly we arrange 
the sample covariance, Poisson variance and total covariance of 
${\MM}$ into matrices ${\Covs}$, ${\Covm}$ and 
${\Cov}={\Covs}+{\Covm}$.    

Given a set of parameters $\theta_\alpha$, the Fisher matrix quantifies the 
information in both the cluster counts and cluster covariance as \cite{LimHu05,LimHu07}
\eqn{
F_{\al\be}=
\del{\MM}{\al}\Covi\del{\MM}{\be}^T
+\frac{1}{2}\Tr{\Covi\del{\Covs}{\al}\Covi\del{\Covs}{\be}}
\,,
\label{eq:fisher}
}
where the first term contains information on the counts and the second 
term the information on the covariance of these counts. The clustering properties of galaxy clusters - encoded in their covariance - bring extra information to the cluster counts, which helps on self-calibration 
\fixed{(constraining {\it nuisance} parameters along with the cosmology)}
of observable-mass distribution \cite{MajMoh04,LimHu05,LimHu07,Gladders07,Baxter16} and, as we shall see, of the cluster selection function. 
The inverse Fisher matrix approximates the covariance matrix of the parameters 
$C_{\alpha \beta}\approx [ F^{-1}]_{\alpha \beta}$. The marginalized error 
on a single parameter $\theta_\alpha$ is 
$\sigma(\theta_\alpha)=[F^{-1}]_{\alpha \alpha}^{1/2}$.
In case we have {\it prior} information on parameter $\theta_\alpha$  
at the level of $\sigma_p(\theta_\alpha)$, we add to the 
Fisher matrix a diagonal contribution of 
$\sigma_p^{-2}(\theta_\alpha)\delta_{\alpha \beta}$ before inversion. 

Finally, variations on the number counts of $\Delta \MM$ and 
on the sample covariance of $\Delta \Covs$, relative to their 
values in the fiducial model, induce a systematic error or 
bias $b(\theta_\alpha)=\delta \theta_\alpha$ 
on a derived parameter $\theta_\alpha$, given by
\cite{Bern09,Eri11}
\eqn{
b(\theta_\al) = 
  F^{-1}_{\al\be}\left\{
\del{\MM}{\be} \Covi \Delta \MM 
+ \frac{1}{2}\Tr{\Covi\del{\Covs}{\be}\Covi\Delta{\Covs}}
\right\} \,. \nonumber \\
\label{eq:fbias}
}

This equation can be used for assessing the bias on inferred 
cosmological parameters when neglecting the inclusion of 
selection function parameters, given that the true counts 
in the fiducial model require these additional parameters. 

\section{Fiducial Model} \label{sec:fiducial}

\begin{figure*}
\includegraphics[scale=1]{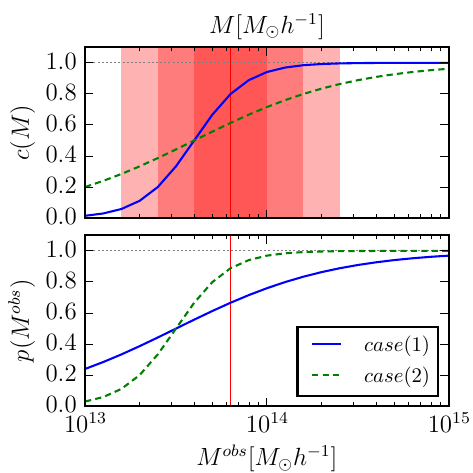}
\includegraphics[scale=1,page=3]{\select/paper_cpvals.pdf}
\caption{
Completeness and purity as a function of mass for cases (1) and (2) 
at $z=0$.
The red vertical line denotes the threshold mass $M_{th}^{\rm obs}=10^{13.8}\Msun/h$ assumed in the fiducial model.
({\it Left}): both functions are shown separately and the red shaded regions display the mass spread around this threshold at 1, 2 and 3$\sigma_{\ln M}$.
({\it Right}): Ratio of completeness and purity (see Eq. \ref{eq:cef}) 
as a function of mass. As a result, case (1) produces an increase on 
cluster counts for higher masses and a decrease at lower masses, while case (2) induces the opposite behavior.
}
\label{fig:cp_form}
\end{figure*}

We choose a fiducial cosmology from a flat $w$CDM model with 
 best-fit parameters consistent with the results form Planck \cite{Planck15}, as $h^2\Omegam=0.14$, $h^2\Omega_b=0.022$, $w=-1$, $A_s=2.13\times10^{-9}$ (corresponding to $\sigma_8=0.83)$, 
$n_s=0.96$, $\tau=0.089$. 
We also set priors of 1\%  on all 
parameters, except for the $h^2\Omegam$ and $w$, which will vary 
freely as we wish to study the potential for galaxy clusters to constrain 
dark energy in the presence of cluster 
selection parameters.  


We assume a survey area of $5000$ deg$^2$, similar to that planned 
for the final observations of the Dark Energy Survey (DES) \cite{DES05}. We consider the counts and 
covariance within $500$ cells of $10$ deg$^2$ each.
To reflect expectations and limitations of cluster finders in current 
optical surveys, we restrict the analysis to $9$ redshift bins of $\Delta z=0.1$ 
from $z=0.1$ to $\zm=1.0$.
We also include seven bins of observed mass of $\Delta \logMobs{}=0.2$ 
from a threshold mass of $M_{th}^{\rm obs}=10^{\mth}\Msun/h$,
where the last bin was reshaped to $\logMobs{}=[15.0:17.0]$ to include all 
high-mass clusters. 
\fixed{This binning choice leaves us with 63 bins of count measurements,
which we expect to be sufficient for providing information on the observed mass and redshift evolution of the selection function and the observable-mass relation.
The completeness and purity mostly shift each bin individually
(although to consider the full effect of completeness within an observed mass bin it is necessary to integrate over all masses, see Fig.~\ref{fig:cp_form}).
The scatter on the observable-mass relation spreads a portion of the clusters across different mass bins,
and the mass bias systematically shifts clusters to higher (or lower) observed mass bins.
This approach also allows us to test the cluster constraining power when considering different minimum masses, by simply adding (or removing) mass bins.
}

The observable-mass $P(\Mobs|M)$ distribution will be assumed to be Gaussian in 
$\ln M$ with a scatter $\sigma_{\ln M}$ and bias $\ln M_{bias}$ 
\bea
P(\Mobs|M)=
\frac{1}{\sqrt{2\pi\sigma_{\ln M}^2}}
\exp\cuad{-\frac{\chi^2\de{\Mobs}}{2}}
\eea

\noindent
where
\bea
\chi\de{\Mobs}=
\frac{\ln\Mobs-\ln M -\ln M_{bias}}{\sigma_{\ln M}}.
\label{eq:chi}
\eea

\fixed{We parametrize the evolution of the mass bias with redshift as \cite{LimHu05},}
\bea
\ln M_{bias}(z)=A_b+n_b\ln(1+z),
\eea

\noindent
where the fiducial values are $A_b=n_b=0$.
Since we expect the mass scatter in the relation to increase for high redshifts and low masses, 
we take 
\bea
\frac{\sigma_{\ln M}^2(z,M)}{0.2^2}=1 + B_0 + B_z (1+z) \\ + B_M \de{\frac{\ln M_{s}}{\ln M}},\nonumber
\eea
with the fiducial values of $B_0=B_z=B_M=0$ and we fix the pivot mass $M_s=10^{14.2}M_\odot/h$ .

As clusters of high mass stand out in observations, 
we expect less ambiguity in detecting them. Therefore the 
completeness and purity should approach unity at high enough values of 
$M$ and $\Mobs$. 
Similarly, for low masses, the number of clusters increase and we expect 
the confusion to be larger, so the completeness and purity decrease. 
We set a functional form for both completeness and purity that interpolates 
between these two limits of high and low masses as
\newcommand{\xc}{\mathcal{M}_c(M, z)}
\newcommand{\xp}{\mathcal{M}_p(\Mobs, \zphot)}
\newcommand{\nc}{{n_c}}
\newcommand{\np}{{n_p}}
\newcommand{\Mc}{[M/M_c(z)]^\nc}
\newcommand{\Mp}{[\rich/\rich_p(z)]^\np}

\bea
c(M, z)&=&\frac{\Mc}{\Mc+1}\,, \\
p(\rich, \zphot)&=&\frac{\Mp}{\Mp+1}\,,
\eea
where $M_c(z)$ and $M_p(z)$ are parametrized functions and 
we take the exponents $\nc$ and $\np$ to be constants. 
\fixed{
This functional form was characterized and shown to describe well completeness and purity in simulations \cite{Aguena_etal_inprep}.
}
We consider 2 different cases, as shown in Table~\ref{tab:cp}:
case (1) sets values $\nc=3$ and $\np=1$, therefore the ratio $c/p$ goes to zero in the limit of low $M$ and $\rich$;
case (2) sets values $\nc=1$ and $\np=3$, therefore the ratio $c/p$ goes to infinity in the limit of low $M$ and $\rich$.
These two cases should bracket a reasonable range of possible parametrizations for 
the selection and their dependence on mass and redshift. 
For the mass scales $M_c$ and $\Mobs_p$, which control 
the transition in completeness and purity function, we take linear relations:
\bea
\log M_c(z)&=&\log \tilde{M}_{c}+c_0+c_1(1+z) \\
\log \Mobs_p(z)&=& \log \tilde{M}^{\rm obs}_{p}+p_0+p_1(1+z)
\eea
with fiducial values of $c_0=p_0=c_1=p_1=0$.  
Here $\tilde{M}_{c}$ and $\tilde{M}^{\rm obs}_{p}$ are arbitrary 
pivot masses where completeness and purity decrease to $50\%$ in the fiducial model. 
For illustrative purposes we fix them to $\tilde{M}_{c}=10^{13.6}\Msun/h$ 
and $\tilde{M}^{\rm obs}_{p}=10^{13.5}\Msun/h$, 
which results in completeness $\approx (80\%,61\%)$
and purity $\approx (67\%,89\%)$ around the threshold mass for cases (1,2).

For reference, we consider an additional case of perfect cluster detection, i.e. completeness and purity
equal to unit for all masses and redshifts.
We will denote this as case (0) and will consider the bias induced 
on dark energy parameters when case (0) is assumed whereas 
the true model is either case (1) or (2). We will also consider 
the dark energy constraints derived within cases (1) and (2) 
and the impact of prior knowledge on {\it nuisance} parameters describing the 
observable-mass relation and cluster selection. 

The functional forms proposed for completeness and purity are shown on the left panel of Fig.~\ref{fig:cp_form}.
While purity is a function of the observed mass of clusters,
completeness depends on true mass of the dark matter halos.
Therefore for a given value of observed mass,
the effective completeness results from the contribution of a range of true masses determined by the scatter in the observable-mass relation.
This feature is illustrated on the left panel of Fig.~\ref{fig:cp_form},
where the vertical red line indicates the fiducial observed mass threshold $\Mobs_{th}=10^{13.8}\Msun h^{-1}$),
and the red shaded regions delineate the scatter at 1, 2 and 3$\sigma_{\ln M}$ levels for the effective selection.

The right panel of Fig.~\ref{fig:cp_form} shows the ratio of completeness and purity ($c/p$), 
which affects the effective cluster selection in Eq.~\ref{eq:cef}.
For each of the cases (1) and (2),
the ratio $c/p$ has limits indicated in Table~\ref{tab:cp}.
In both cases, the ratio $c/p\rightarrow 1$ in the limit of high masses, since both $c$ and $p$ approach unit in this limit.  
For case (1) the ratio $c/p\rightarrow0$ at lower masses;
however, in the mass range investigated ($\geq 10^{13.8}\Msun h^{-1}$),
the ratio $c/p>1$, resulting in more detected clusters than case (0). 
An opposite effect occurs for case (2),
resulting in fewer cluster detections.
\fixed{Although the region of very low masses ($\rich\sim10^{13}\Msun h^{-1}$) is not the focus of this work,
it is interesting to analyze what happens to the ration $c/p$ in this limit.
For case (1), the completeness decreases faster than the purity,
meaning that the capability of the cluster finder detect objects goes to zero.
The other case,
where $c/p$ becomes large at low masses,
happens when purity decreases faster than completeness. 
This may happen for instance as the cluster finder attempts to detect clusters whose BCG has magnitude close to the survey limiting magnitude,
particularly at low masses and high redshifts.
As the cluster finder struggles with the detection,
the number of false positives may become larger than the number of missed clusters.}

\begin{table}
\begin{tabular}{c*{4}{c}rr}
\hline
Case & completeness	& purity & $c/p ( \text{ as } M\rightarrow 0$)\\
\hline
0  	& $c=1$ & $p=1$ &  1	\\
1  	& $n_c$=3	& $n_p$=1  & 0	\\
2  	& $n_c$=1	& $n_p$=3  & $\infty$	\\
\end{tabular}
\caption{Cases considered for completeness and purity parameter values.}
\label{tab:cp}
\end{table}

\fixed{Notice that by using multiple observables, we attempt to self-calibrate various 
nuisance parameters describing the observable-mass relation and the selection effects. 
In reality, this will only be effective if the parametrizations used are indeed correct.}

\section{Results} \label{sec:results}

\begin{figure*}
\includegraphics[scale=0.75]{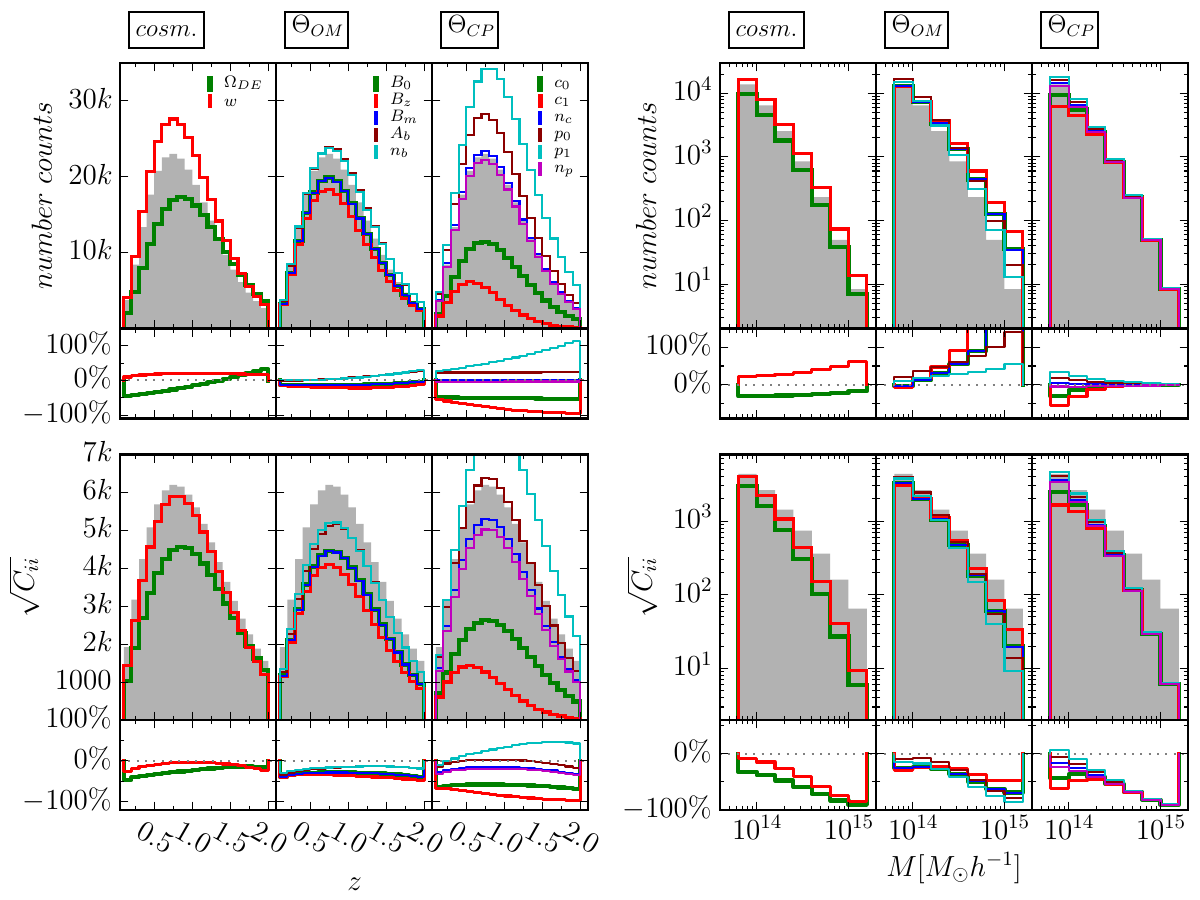}
\caption{
\fixed{Variation of the cluster number counts ({\it top}) and sample variance ({\it bottom}) as a function of redshift ({\it left}) and mass ({\it right}) 
for changes in parameters of dark energy, observable-mass relation and completeness and purity, considering case (1).
The left panels were computed for the mass bin $\Mobs_{th}=[13.8:14.0]$ and the right panels for the redshift bin $z=[0.6, 0.7]$.
}
The \fixed{gray shaded region} is the fiducial case,
and the colored lines indicate a variation of $+0.2$ 
in each parameter.
}
\label{fig:nc}
\end{figure*}

Before studying the impact of selection parameters on dark energy constraints, for illustrative purposes we first look at the effect on cluster 
abundance and clustering from each cosmological and {\it nuisance} parameter.
\fixed{
Fig.~\ref{fig:nc} shows the effects of cosmology and selection on the number counts (top) and the diagonal of sample covariance (bottom)
as a function of redshift (left) and mass (right),
with selection parameters from case (1).
The left panels, where different redshift bins are displayed,
were computed in the mass bin $\Mobs_{th}=[13.8:14.0]$,
and the right panels in the redshift bin $z=[0.6, 0.7]$,
as those are the bins with the largest number of objects,
having a more significant contribution on the constraints.
We note that the actual constraining results make use of all mass and redshift bins and the non-diagonal terms of the sample covariance.
}
We compute counts for the fiducial model (\fixed{gray shaded region}) and for positive variations of $0.2$ in each parameter considered. 

\fixed{
To evaluate some of the effects the parameters considered have on the constraints,
let us explore the changes on the number counts as a function of redshift (top left panel of Fig.~\ref{fig:nc}).
}
We assume a flat universe, so an increase in $\Omega_{\rm DE}$ results in a reduction for $\Omega_m$ and the overall abundance of clusters is reduced.
Increasing $w$ causes the dark energy behavior to 
be closer to that of non-relativistic matter,
also resulting in an increase on cluster counts.

From the definition of mass bias $M_{\rm bias}$ in Eq.~\ref{eq:chi},
increasing its value results in a lower effective mass threshold, therefore increasing the counts of clusters.
The same is true for the mass scatter \cite{LimHu05}, though with a lower sensitivity compared to the mass bias.

Increasing the completeness parameters $c_0$ and $c_1$
increases the mass scale $M_c$ in which completeness becomes $50\%$, lowering the values of completeness across all masses
and reducing the counts of detected clusters.
An increase on $n_c$ makes the drop in completeness at $M<M_c$ sharper,
resulting in a slight increase of completeness for $M>M_c$ and a decrease for $M<M_c$.
Since the mass threshold adopted ($\Mobs_{th}=10^{13.8}\Msun h^{-1}$) is higher than the fiducial value of $M_c$ ($10^{13.5}\Msun h^{-1}$),
increasing $n_c$ produces a slight increase in the counts.

Finally, given our effective cluster selection from Eq.~\ref{eq:cef},
purity has an inverse effect compared to the the completeness for the the counts. In fact, since completeness and purity have the same functional form,
changes in each purity parameter causes opposite effects on counts compared to changes in the corresponding completeness parameter.

These results indicate how parameters are (anti) correlated, i.e. how changes in one parameter can compensate for changes in other parameters.
These effects, however, reflect the dependency around 
the fiducial model when fixing all other parameters at their 
fiducial value. When marginalizing over parameters,
the resulting correlations may change.

\subsection{Selecting Cases} \label{sec:select}
The first issue we consider is whether it is worth including completeness and purity parameters in the cluster analysis for purposes of constraining dark energy.
Including extra {\it nuisance} parameters (\fixed{related to the observational effects}) increases the accuracy,
but decreases the precision of cosmological constraints.
When completeness and purity effects are ignored, i.e. when case (0) is assumed despite imperfect selection,
the resulting cosmological parameters $\theta_\al$ constrained have a bias $b(\theta_\al)$ (Eq.~\ref{eq:fbias}).
The assumption of perfect detection can still provide reliable cosmological 
parameter constraints as long as the bias is smaller than the 
parameter constraints
\bea
b(\theta_\al) \lesssim \gamma
\sigma(\theta_\alpha)=
\gamma \de{F^{-1}}^{1/2}_{\al\al}\,,
\label{eq:validation}
\eea
\noindent
where $\gamma=1,2,3$ indicate biased predictions inside the $68, 95, 99\%$ confidence levels. 
Here 
$\Delta \MM$ and $\Delta \Covs$ in Eq.~\ref{eq:fbias} are the differences in counts 
and sample covariance between 
predictions in case (0) and cases (1,2).

\begin{figure}[!h]
\includegraphics[scale=1,page=3]{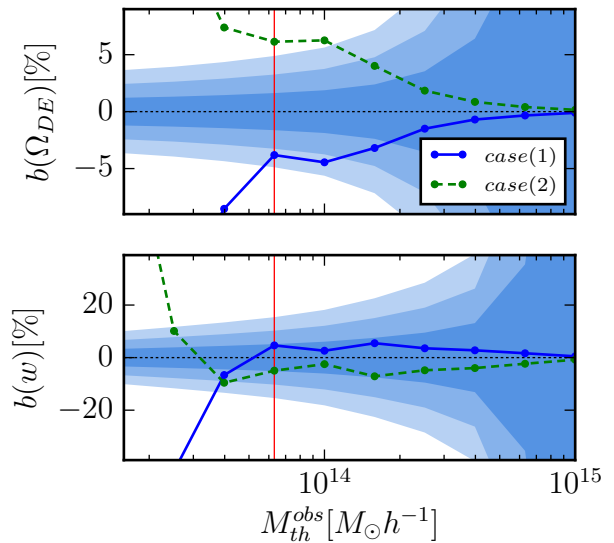}
\caption{Comparison between i) the 1, 2, 3$\sigma$ constraints (blue shaded regions) on dark energy parameters $\theta_{\rm DE}=(\OmegaDE,w)$ for case (0) of perfect cluster selection and the ii) percent bias $b(\theta_{\rm DE})$ (lines) on dark energy caused by ignoring completeness and purity effects as given by cases (1,2) (solid, dashed).
As the bias becomes comparable to 1$\sigma$ constraints $b(\theta_\al) \approx \sqrt{F^{-1}_{\al\al}}$,
the assumption of perfect detection results in significantly incorrect best-fit predictions.
For the threshold mass considered of $\logMobs{th}=13.8$ (vertical red line),
the bias $b(\OmegaDE)$ is larger than the corresponding 2$\sigma$ constraint for both cases (1,2), whereas  $b(w)$ is comparable to the 1$\sigma$ constraint. 
}
\label{fig:bias}
\end{figure}

Fig.~\ref{fig:bias} shows the bias induced on dark energy parameters $(\OmegaDE, w)$ as a function of the observed mass threshold used $\Mobs_{th}$, if we assume case (0) when in reality counts are described by case (1) (solid line) and (2) (dashed line). Also shown are the 
1,2 and $3\sigma$ confidence levels on $(\OmegaDE, w)$ in case (0) (blue shaded regions).
\fixed{
This observed mass threshold we consider does not imply that only a single bin of mass is being used,
but that mass bins of $\Delta \logMobs{}=0.2$ down to this threshold are being used.
Therefore, $\logMobs{th}=13.8$ results in 7 observable mass bins,
while $\logMobs{th}=14.2$ considers only 5 observable mass bins.}
The bias on $\OmegaDE$ 
\fixed{has surpassed the 1$\sigma$ constraints for thresholds $\logMobs{th}\leq14.2$}
in both cases (1) and (2). 
In fact, the bias is larger than 2$\sigma$ for case (1) and 3$\sigma$ for case (2) around the fiducial threshold mass $\logMobs{th}=13.8$.
The bias on $w$ is around 1$\sigma$ at $\logMobs{th}=13.8$, indicating that this parameter is less sensitive to the selection effects. However $w$ is less well constrained than $\OmegaDE$ so a bias comparable to 1$\sigma$ constraints may be even more significant when constraining models of dark energy. 

Notice that the bias behavior as a function of $\Mobs_{th}$ is not monotonic.
This occurs mainly due to the fact that the ratio $c/p$ of  completeness and purity is also not monotonic, as seen on the right panel of Fig.~\ref{fig:cp_form}.
For very large thresholds $c/p$ indeed approaches unit, as in case (0) and the bias is small. For masses around the fiducial threshold, the bias is caused mainly by the upper/lower bump in $c/p$ for case (1)/(2). For lower masses, the bias becomes much larger and is dominated by the rapid change in $c/p$ for both cases (1) and (2).

\begin{figure*}
\includegraphics[scale=1]{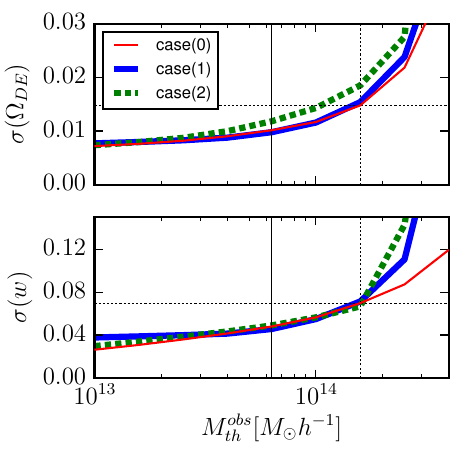}
\includegraphics[scale=1]{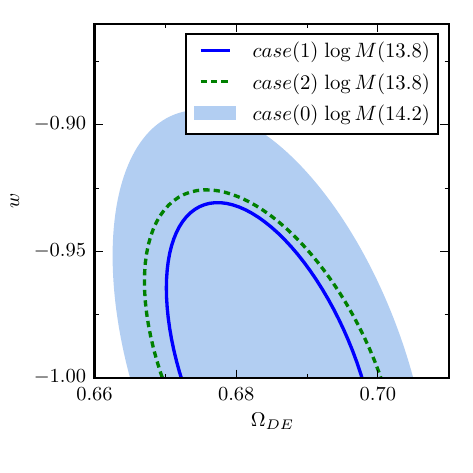}
\caption{({\it Left}): Constraints on dark energy 
parameters $(\OmegaDE, w)$ as a function of threshold 
mass for different cases (0), (1) and (2). 
Even though the constraints are somewhat similar, 
for case (0) they are only reliable down to $\logMobs{th}=14.2$ (dotted vertical line). 
({\it Right}): Constraints for cases 
(0),(1) and (2) at different threshold masses. 
The blue shaded region shows constraints for case (0) 
under its minimum threshold $\logMobs{th}=14.2$.
Both cases (1) and (2), which include completeness and purity and go to a lower threshold $\logMobs{th}=13.8$, produce better constraints than case (0).
}
\label{fig:ell_comp}
\end{figure*}

\fixed{
Given that the minimum mass threshold that still allows for somewhat reliable dark energy constraints under case (0) is located at $14.2<\logMobs{th}<14.4$, we now investigate for what mass thresholds the constraints under cases (1) and (2) become better than those from case (0) under 
$\logMobs{th}=14.2$ as a conservative comparison.
}
As we go to lower threshold masses 
and need to fully model the selection with a larger 
number of {\it nuisance} parameters (\fixed{describing completeness and purity}), we also increase considerably the number of clusters probed, 
which brings more cosmological information. 

The left panel of Fig.~\ref{fig:ell_comp} shows 1$\sigma$ constraints for cases (0), (1) and (2) as a function of the 
observed mass threshold $\Mobs_{th}$.
The dotted lines mark the mass threshold $\logMobs{th}=14.2$ and the corresponding constraints for case (0). 
As we decrease the threshold mass,
the constraints for cases (1,2) improve.
At the fiducial threshold $\logMobs{th}=13.8$,
the marginalized constraints of $\OmegaDE$ and $w$ 
for both cases (1) and (2) are lower than 
those from case (0) with threshold $\logMobs{th}=14.2$.

The right panel of Fig.~\ref{fig:ell_comp} shows the joint dark energy constraints for multiple cases at different 
thresholds.
The solid and dashed lines correspond to cases (1) and (2) respectively with the fiducial threshold, whereas the blue shaded region corresponds to case (0) and the higher threshold in which this case is marginally reliable.
We see that a fiducial threshold $\logMobs{th}=13.8$ is enough to significantly improve dark energy constraints relative to case (0), despite the increase in the number of {\it nuisance} parameters from the selection function.

It is interesting to notice that, as we consider even lower threshold masses than the fiducial one assumed here, we continue to improve dark energy constraints. However that requires us to trust that the selection can still be well described by the parametrized functional forms assumed here down to those lower masses. That assumption has to be backed up by multiple methods, including trustworthy simulations and comparisons to other cluster detections at multiple wavelengths. Using a slightly incorrect selection at low masses could highly bias the derived constraints. 
More conservatively, 
in going to lower masses, one needs to consider more general forms for the selection with increasing number of {\it nuisance} parameters (\fixed{modeling the selection function and mass-richness relation}), which would likely degrade  cosmological constraints. 

It becomes 
clear nonetheless that if one can properly model 
the survey completeness and purity down 
to levels of around $\sim 60\%$ -- for which the assumption of perfect selection can no longer be made -- the information in cluster counts and clustering is enough to self-calibrate
observable-mass and selection parameters
\fixed{(constrain the parameters along the cosmology)}
,
providing better dark energy constraints than fixing 
conservatively higher thresholds in order to ignore 
selection effects.

\subsection{Completeness and Purity Effects} \label{sec:cpeff}
\begin{table}
\setlength{\tabcolsep}{6pt}
\begin{tabular}{cc|cc|cc{c}rr}
\hline
{\color{white}case (0)} && Case (1) && Case (2) \\ 
$\theta_{\rm OM}$ & $\theta_{\rm CP}$ & $\sigma(\OmegaDE)$ & $\sigma(w)$ & $\sigma(\OmegaDE)$ & $\sigma(w)$\\
\hline
\hline
fix	& fix		& \sigxxOmB	& \sigxxwB	& \sigxxOmC	& \sigxxwC	\\
free	& fix		& \sigfxOmB	& \sigfxwB	& \sigfxOmC	& \sigfxwC	\\
fix	& free		& \sigxfOmB	& \sigxfwB	& \sigxfOmC	& \sigxfwC	\\
free	& free		& \sigffOmB	& \sigffwB	& \sigffOmC	& \sigffwC	\\
$\prv\%$& free		& \sigpfOmB	& \sigpfwB	& \sigpfOmC	& \sigpfwC	\\
free	& $\prv\%$	& \sigfpOmB	& \sigfpwB	& \sigfpOmC	& \sigfpwC	\\
$\prv\%$& $\prv\%$	& \sigppOmB	& \sigppwB	& \sigppOmC	& \sigppwC	\\
\end{tabular}
\caption{Constraints on dark energy parameters $(\OmegaDE, w)$ for different prior on observable-mass 
parameters $\theta_{\rm OM}$ and completeness/purity parameters $\theta_{\rm CP}$.}
\label{tab:cons_cp}
\end{table}

\newcommand{\Deg}{\mathcal{D}}

In this section, for illustrative purposes we focus our discussion on the constraints from case (1), but the results and conclustions for case (2) are similar (see e.g. Table~\ref{tab:cons_cp}).
We start considering baseline constraints for the fiducial model described in \S~\ref{sec:fiducial}, 
assuming perfect knowledge of the observable-mass relation as well as the completeness and purity. 
In this case the dark energy constraints are $\sigma(\OmegaDE, w) = ${\sigfixfix}.
If we let the observable-mass parameters vary freely, but keep the completeness/purity 
parameters fixed, these constraints degrade to {\sigfrefix}. 

\begin{figure}[!h]
\includegraphics[scale=1]{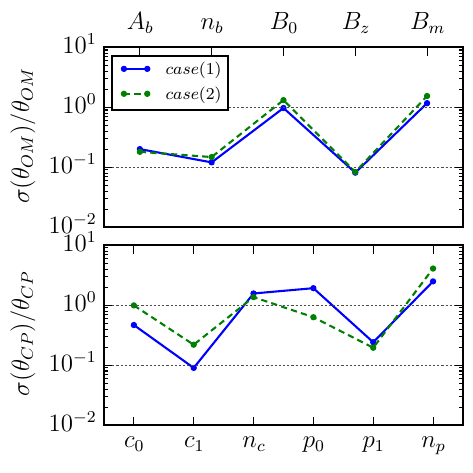}
\caption{Fisher constraints 
derived for {\it nuisance} parameters in case (1) (blue solid line) and case (2) (green dashed line). The parameters are related to the observable-mass relation ({\it top panel}) and completeness/purity ({\it bottom panel}). No priors were 
assumed for these {\it nuisance} 
parameters. 
}
\label{fig:c_npar}
\end{figure}

\begin{figure*}
\begin{center}
\includegraphics[scale=.75,page=1]{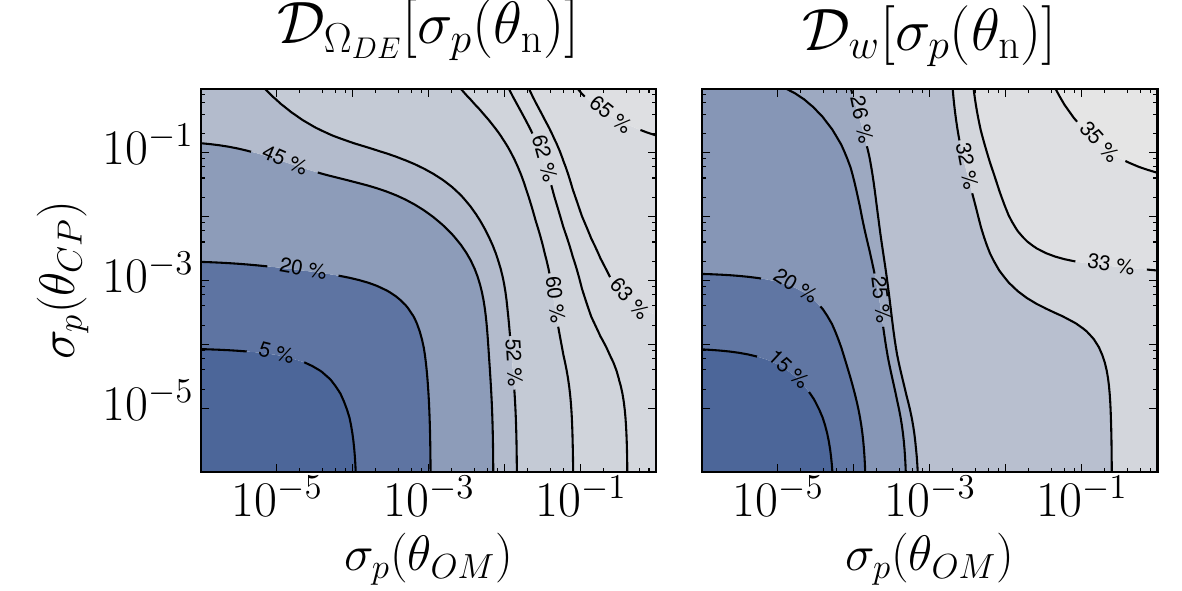}{\centering}
\end{center}
\caption{
Contours of constant degradation ${\cal D_{\theta_{\rm DE}}}$ on constraints of dark energy parameters $\theta_{\rm DE}=(\OmegaDE,w)$ 
as a function of priors $\sigma_{p}(\theta_{\rm OM})$ on all observable-mass relation parameters 
and priors $\sigma_{p}(\theta_{\rm CP})$ on all completeness/purity parameters. The degradation is 
considered for case (1) and relative to the case of perfect {\it nuisance} 
parameters (for which 
$\sigma_{p}(\theta_{\rm OM})=\sigma_{p}(\theta_{\rm CP})=0)$. 
For ${\cal D}_{\theta_{\rm DE}}<20\%$, 
subpercent level priors on all {\it nuisance} parameters are required.
}
\label{fig:priors}
\end{figure*}

Next, we consider the effect of varying the parameters of completeness and purity. 
First we fix the observable-mass parameters and let completeness/purity parameters vary freely. 
In this case the dark energy constraints become {\sigfixfre}.  
If we now let both the observable-mass and completeness/purity parameters vary 
freely, the constraints become {\sigfrefre}. This corresponds to a degradation 
of {\degfrefre} relative to the case where these functions are perfectly known, but of only {\degfrefrefx} relative to 
the case where only the selection if fixed. 
Therefore, including completeness and purity effects on top of observable-mass parameters avoids biased parameters without degrading the constraints significantly.

Finally, in order to quantify the effects of priors $\sigma_{p}(\theta_{\rm n})$ assumed on {\it nuisance} parameters
$\theta_{\rm n}=(\theta_{\rm OM}, \theta_{\rm CP})$, namely observable-mass parameters $\theta_{\rm OM}$ and/or 
completeness/purity parameters $\theta_{\rm CP}$, we define the degradation factor $\Deg_{\theta_{\rm DE}}$ on the constraints of dark energy parameters $\theta_{\rm DE}=(\OmegaDE, w)$
as 
\bea
\Deg_{\theta_{\rm DE}}[\sigma_{p}(\theta_{\rm n})]=
\frac{\sigma[\theta_{\rm DE}| \sigma_{p}(\theta_{\rm OM}), \sigma_{p}(\theta_{\rm CP})]}{\sigma(\theta_{\rm DE})|_{\rm ref}}-1\,.
\label{eq:deg}
\eea

This factor represents the relative difference between 
constraints on $\theta_{\rm DE}$ given priors $\sigma_{p}(\theta_{OM})$ and $\sigma_{p}(\theta_{CP})$ and 
the reference ideal case $\sigma(\theta_{\rm DE})|_{\rm ref}=\sigma[\theta_{\rm DE}| 0,0]$ where 
{\it nuisance} parameters are perfectly known.

Applying priors of $\prv\%$ (or 
$10^{-2}$ when the fiducial value is zero) on the 
observable-mass relation parameters but letting the completeness/purity parameters vary freely, 
the constraints become \sigprifre. Conversely, if we let the observable-mass relation vary 
freely and apply a \prv\% prior on the completeness/purity parameters, the constraints 
become {\sigfrepri}.
Finally, applying a \prv\% prior to all {\it nuisance} parameters (\fixed{related to the observational effects}), 
the constraints become {\sigpripri}. This corresponds to a degradation of 
{\degpripri} relative to the case in which these {\it nuisance} parameters are perfectly known. 

External priors may come from multiple sources, including detailed simulations, lensing masses for a subsample of clusters, or cross-matches to clusters detected at other \fixed{wavelengths},
 e.g. X-ray and/or millimeter. 
In all cases, these priors are likely to provide clues on the correct functional forms 
for these functions and conservative ranges for both the 
observable-mass relation and the completeness/purity parameters.

\newcommand{\prBS}{{$Pr_{OM}$}}
\newcommand{\prCP}{{$Pr_{CP}$}}

Fig.~\ref{fig:c_npar} shows Fisher constraints -- relative to the fiducial value -- 
for each {\it nuisance} parameter $\theta_{n}$.
Given that none of these parameters are constrained to better than 10\%, having 1\% priors on any of these {nuisance} parameters would have an important effect in constraining the parameters themselves. However, as we have seen the effect on improving dark energy constraints is 
very small. 

Fig.~\ref{fig:priors} shows contours of constant degradation ${\cal D_{\theta_{\rm DE}}}$ on dark energy parameters $\theta_{\rm DE}=(\OmegaDE,w)$ -- relative to perfect {\it nuisance} parameters -- as a function of priors on observable-mass relation $\sigma_{p}(\theta_{\rm OM})$ 
and on completeness/purity parameters 
$\sigma_{p}(\theta_{\rm CP})$.
Notice that both panels of Fig.~\ref{fig:priors} present a similar qualitative behavior, though constraints on $w$ 
do not degrade as much as constraints on $\OmegaDE$. 
We see that it is important to improve priors on both observable-mass as well as completeness/purity parameters. For degradations on dark energy 
constraints to remain lower than 20\%, it is necessary 
to have quite strong external priors at the subpercent level, which are clearly very hard to achieve even in optimistic scenarios.

\subsection{Future Surveys} \label{sec:future}

\begin{table}
\setlength{\tabcolsep}{6pt}
\begin{tabular}{c|cc|cc{c}rr}
\hline
{\color{white}case (0)} & \multicolumn{2}{c|}{Case (1)} & \multicolumn{2}{c}{Case (2)} \\ 
$z_{max}$ & $\sigma(\OmegaDE)$ & $\sigma(w)$ & $\sigma(\OmegaDE)$ & $\sigma(w)$\\
\hline
0.3	& \sigOmBIII	& \sigwBIII	& \sigOmCIII	& \sigwCIII	\\
0.5	& \sigOmBV	& \sigwBV	& \sigOmCV	& \sigwCV	\\
0.7	& \sigOmBVII	& \sigwBVII	& \sigOmCVII	& \sigwCVII	\\
1.0	& \sigOmBX	& \sigwBX	& \sigOmCX	& \sigwCX	\\
1.2	& \sigOmBXII	& \sigwBXII	& \sigOmCXII	& \sigwCXII	\\
1.5	& \sigOmBXV	& \sigwBXV	& \sigOmCXV	& \sigwCXV	\\
1.7	& \sigOmBXVII	& \sigwBXVII	& \sigOmCXVII	& \sigwCXVII	\\
2.0	& \sigOmBXX	& \sigwBXX	& \sigOmCXX	& \sigwCXX	\\
\hline
\end{tabular}
\caption{Constraints for dark energy as a function of maximum redshift $\zm$. Here all {\it nuisance} parameters describing the effective selection function (observable-mass, completeness and purity) vary freely.}
\label{tab:cons_z}
\end{table}

\begin{figure}[h]
\includegraphics[scale=1,page=1]{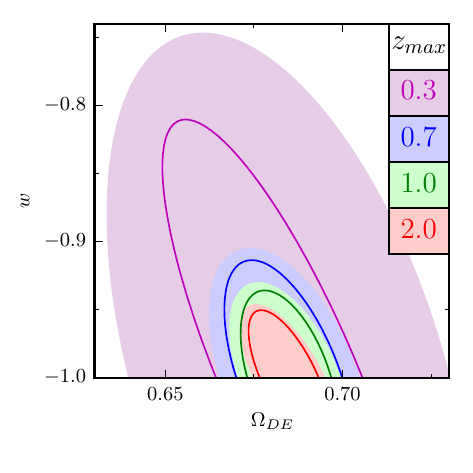}
\caption{Effect on dark energy constraints when changing the survey maximum redshift $z_{max}$ 
from 0.3 (pink), 0.7 (blue), 1.0 (green) and 2.0 (red). Solid lines refer to case (1) and shaded regions refer to case (2).
}
\label{fig:zrange}
\end{figure}

Future surveys \cite{eROSITA,Euclid,LSST},
will allow for improvements on both total survey area and depth and the effects of completeness and purity across 
these improvements tend to become more important. 
The impact of survey depth or maximum redshift $z_{max}$ on dark energy constraints in shown in Fig.~\ref{fig:zrange} and Table~\ref{tab:cons_z}.

Optical cluster finders applied to the SDSS in the last decade were limited to relatively shallow magnitudes. For instance the MaxBCG cluster catalog \cite{MaxBCG} had $\zm=0.3$, which in our Fisher analysis produces constraints $\sigma(\OmegaDE, w) = ${\sigIII},
corresponding to a degradation of {\degIII} relative to our fiducial case ($\zm=1.0$).
More recently, the redMaPPer cluster finder \cite{ RedMapper14,Rykoff16} has been applied to both the SDSS and the DES Science Verification data, producing catalogs that go up to $\zm \sim 0.7$, corresponding to constraints of {\sigVII},
a degradation of {\degVII} relative to our fiducial model. 
Since redMaPPer makes use of the red sequence for 
detecting optical clusters, it may be challenging to extend 
its results to redshifts much larger than these. 

For upcoming surveys planned to extend observations to higher redshifts,
we find constraints of {\sigXX} for $\zm=2$,
an improvement of {\degXX}. 
Case (2) presents a higher degradation when lowering $\zm$ than case (1),
however,
the improvement is lower when we extend $\zm$.

\begin{figure}[h]
\includegraphics[scale=1,page=1]{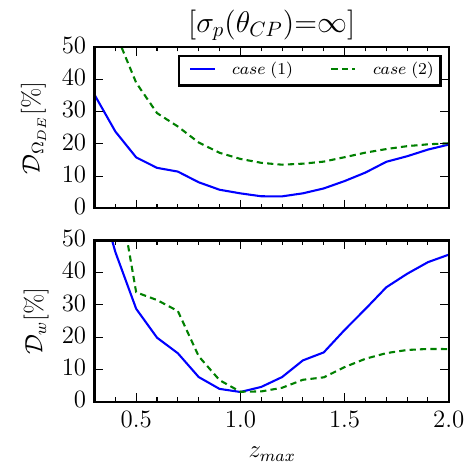}
\caption{Percent degradation ${\cal D}_{\theta_{\rm DE}}$ on constraints for dark energy parameters $\theta_{\rm DE}=(\OmegaDE, w)$ as a 
function of maximum redshift $z_{max}$, for selection 
function parametrized in case (1) (solid line) and case (2) (dashed line). Degradations are computed for the case where the completeness and purity parameters are free [$\sigma_p(\theta_{\rm CP})=\infty$] relative to the case where these parameters are perfectly known  [$\sigma_p(\theta_{\rm CP})=0$].
${\cal D}_{\theta_{\rm DE}}$ decreases with $\zm$ up to $z_{max} \sim 1$ and increases for higher redshifts. The dark energy constraints themselves always improve for higher values of $\zm$, but the relative sensitivity to 
knowledge on selection parameters increases.
}
\label{fig:zcp}
\end{figure}

We now quantify the impact of completeness and purity for different values of $\zm$ by considering 
the degradation ${\cal D}_{\theta_{\rm DE}}$ 
on dark energy constraints from Eq.~\ref{eq:deg}, 
for the case with free completeness and purity parameters $\sigma[\Omega_{\rm DE}|\sigma_p(\theta_{CP})=\infty]$
relative to the case of 
perfect knowledge $\sigma[\Omega_{\rm DE}|\sigma_p(\theta_{CP})=0]$.
In Fig.~\ref{fig:zcp} we see that 
${\cal D}_{\theta_{\rm DE}}$ 
has a significant overall improvement (i.e. decrease) 
with the increase of $z_{max}$ for cases (1,2),
up to $z_{max}\sim 1.0-1.2$. Beyond those 
redshifts, the degradation increases again, especially 
for $w$ in case (1). Notice however that these 
higher degradations are on top of much improved 
dark energy constraints (see Table~\ref{tab:cons_z}).
Therefore to fully exploit improvements on 
cluster dark energy constraints coming from larger survey depths, it will be important to properly account for 
selection effects, despite the fact that it may be significantly harder to quantify these effects at these higher redshifts.

Finally we quantify the effect of changes in survey 
area. We keep our approach of considering sample covariance from cells of $10$ deg$^2$ and notice that the Fisher Matrix has a linear dependence on total area.
This means all constrained parameters have the same degradation/improvements due changes on the survey area.
Our fiducial area of $5000$ deg$^2$ is similar to what will be observed by the DES.
An area twice as large ($1/4$ of sky) results in an improvement of $\sim29\%$ on both dark energy constraints 
For half-sky is observations, constraints improve by $\sim50\%$, 
and for full-sky they improve by $\sim 65\%$.

\section{Discussion} \label{sec:discussion}

We have explored the effects of completeness and purity on dark energy constraints from the abundance and clustering of galaxy clusters. We parametrized the selection of cluster 
samples to reflect a decrease in completeness and purity 
at lower masses such that they both reach $\sim 50\%$ at a mass scale 
$M\sim 10^{13.5} M_{\odot}/h$. The ratio ($c/p$) determines 
the effective selection. Within our parametrization, 
($c/p$) either goes to zero (case 1) or infinity (case 2) as $M\rightarrow 0$.

We first considered the bias induced on dark energy constraints when neglecting completeness and purity effects from cases (1) and (2). We found that the bias becomes comparable to dark energy 
constraints at a threshold mass of $\Mobs_{th} \sim 10^{14.2} M_{\odot}/h$. As this represents the minimum threshold 
for which it is safe to ignore selection effects, we then 
proceeded to study the inclusion of completeness and purity 
parameters in dark energy constraints for a lower fiducial 
mass threshold of $\Mobs_{th} \sim 10^{13.8} M_{\odot}/h$ .

Since the effective selection includes not only completeness and purity but also the observable-mass distribution, the impact of including completeness and purity depends on assumptions made for the observable-mass parameters. Within 
case (1), baseline constraints for fixed observable-mass parameters and 
fixed completeness and purity are $\sigma(\OmegaDE, w)=(0.006, 0.033)$ and when only completeness and purity parameters vary 
freely, these degrade to $(0.009, 0.042)$. On the other hand, if observable-mass parameters vary freely whereas completeness and purity parameters remain fixed, constraints are $\sigma(\OmegaDE, w)=(0.009, 0.044)$ and they only degrade to $(0.010,0.046)$ 
if completeness and purity also vary freely.

Next we considered the impact of external priors on observable-mass and completeness/purity parameters. From the perspective of dark energy constraints these are {\it nuisance} parameters (\fixed{related to the observational effects}). 
We find that joint priors 
on all {\it nuisance} parameters need to be known to better than $\prv\%$ in order to improve dark energy constraints significantly; with these priors, constraints are restored to $\sigma(\OmegaDE, w)=(0.006, 0.041)$ for case (1).

Although it seems 
unlikely that external priors on selection parameters will reach sub-percent levels for current and upcoming clusters surveys,  interesting priors should be possible from a combination of multiple sources, including detailed simulations, 
cross-matches from other surveys and follow-up spectroscopic observations for a fraction of the cluster sample. For instance, the DES has developed detailed simulations that mimic all its observational properties \cite{Park16, Suchyta16}. 
By running optical cluster finders on these simulations, it is possible to characterize observable-mass and completeness/purity functions \cite{Aguena_etal_inprep}.
Moreover, DES has a significant overlap with the South Pole Telescope (SPT), so cross-matches of DES optical clusters and SPT SZ clusters allow for calibrations on the observable-mass relation \cite{Sar15}. A similar calibration can 
be achieved from X-ray detections \cite{Mehrtens_etal_12,Boller_etal_16} and lensing masses \cite{Melchior16}.

Even though our results indicate that only very stringent (and hard to achieve) priors on {\it nuisance} parameters would be effective in improving dark energy constraints from self-calibrated constraints
\fixed{(where the {\it nuisance} parameters were constrained along the cosmology)},
such priors are actually very important for checking the validity of the assumed functional forms, providing consistency checks for internal self-calibration of {\it nuisance} parameters.

We also investigated the effect of changing 
survey area (from our fiducial $\Delta\Omega=5000$ deg$^2$) and maximum redshift (from fiducial $\zm=1.0$), reflecting expectations from future surveys.
For $\Delta\Omega=10000$ deg$^2$ ($1/4$ of sky) the constraints would improve by $\sim29\%$ and 
for $\Delta\Omega=40,000$ deg$^2$ (full-sky) by $\sim 65\%$, relative to the fiducial case.
If we expand the maximum redshift to $\zm = 2.0$, 
constraints on on $(\OmegaDE,w)$ improve by {\degXX} 
on for case (1), though most of this improvement is already achieved for 
$\zm=1.5$. Despite the improvements on the constraints for higher reshifts and survey areas, 
these constraints also degrade  more significantly
in the lack of knowledge of selection parameters. 
Therefore to fully exploit  the gain in precision, 
it will be even more important to better understand and calibrate the cluster selection function.

Our results were based on the parametrized functions chosen for the effective selection, and 
they may depend to some extent on these choices.
We proposed functional forms for completeness and purity, which are inspired by ongoing 
work involving runs of cluster finders on DES simulations,
\fixed{which we will present elsewhere} 
\cite{Aguena_etal_inprep}.
In fact, our parametrizations bracket a considerable range of possibilities, so we do not expect significant changes in our conclusions when considering alternative parametrizations. On the other hand, when extending cluster analyses to significantly lower mass thresholds, one needs to be assured that the functional forms are still valid down to those masses, which may be hard even with simulations and multi-wavelength cross-matches.
In particular, as $c/p$ become lower than $50\%$, 
we probably need to consider more general functions (or even an arbitrary behavior) for completeness and purity, which may then significantly degrade dark energy constraints (or even bias them for an oversimplified selection), despite the increase in the number of clusters probed. Again, we envision that 
detailed simulations 
and cross-matches should help us in defining 
the most appropriate parametrizations.

\fixedII{
Given that the halo mass-function must be known to high precision for cosmological applications \cite{Cunha10,McCetal18},
and the fact that the Tinker mass-function is only precise at the $5\%$ level \cite{Tin08},
we investigated the effect of changing the halo mass-function prescription in our analysis. In real-data analysis one is expected to make use of well-calibrated fitting formulae or emulators for the mass-function (see e.g. \cite{McCetal18}). 
For illustrative purposes we replaced the Tinker mass-function by the fitting formula from Jenkins \cite{Jenkins01}
We found that this change on the mass function causes variations of up to $20\%$ on the number counts,
which leads to changes $\leq30\%$ on the cosmological constraints and $\leq80\%$ on the constraints of nuisance parameters.
However, the degradation effects on the dark energy constraints caused by the inclusion of nuisance parameters remain comparable to when we use the Tinker mass function (the largest degradation occurs when one of the two sets of nuisance parameters $\theta_{\rm OM}$ or $\theta_{\rm CP}$ is introduced,
and the inclusion of the second set is negligible),
as does the bias on dark energy parameters introduced by ignoring selection function effects.
Therefore, the main conclusions regarding inclusion/exclusion of nuisance parameters are the same as those presented throughout this work.
}

Although intrinsic degeneracies always 
remain to some extent, further improvements in the theoretical modeling of 
cluster properties coming from N-body and gas-dynamics simulations will improve our 
knowledge of the halo mass-function and bias 
in the presence of baryonic effects \cite{Cui12, Bocquet16}, and 
help define appropriate functional forms for the 
observable-mass relation and its intrinsic scatter \cite{Nag06,Kra06,Nag07}. Improvements on semi-analytical Halo Occupation 
Distribution models will also allow for the creation of reliable mock galaxy catalogs 
on which we may run cluster finders and calibrate cluster selection parameters. These theoretical developments combined with external calibrations from cluster cross-matches are essential for cluster cosmology. The self-consistency between observations and theory predictions -- which account for all relevant observational effects -- will advance our knowledge 
of the astrophysical processes that regulate 
observed cluster properties and simultaneously lead to trustworthy cluster cosmological constraints.

\acknowledgements

We thank Christophe Benoist, Vinicius Busti, Ricardo Ogando and Luiz da Costa for useful discussions.
MA is supported by FAPESP. ML is partially supported by FAPESP and CNPq. 
\bibliography{ClustersFisher}

\end{document}